\documentclass[]{article}

\begin{document}

%\scriptsize
\title{Continuous-time histories: observables, probabilities,  phase space
structure and the classical limit}
\author{Charis Anastopoulos \thanks{charis@physics.umd.edu} \\
Department of Physics, University of Maryland,\\
 College Park, MD20742, USA}
\maketitle

\begin{abstract}
The continuous-time histories programme stems from the consistent
histories approach to quantum theory and aims to provide a fully
covariant formalism for quantum mechanics. In this paper we
examine some structural points of the formalism. We demonstrate a
general construction of history Hilbert spaces and identify a
large class of time-averaged observables. We pay particular
attention to the construction of the decoherence functional (the
object that encodes probability information) in the
continuous-time limit and its relation to the temporal structure
of the theory. Phase space observables are introduced, through the
study of  general representations of the history group, which is
the analogue of the canonical group in the formalism.  We can also
define a closed-time-path (CTP) generating functional for each
observable, which encodes the information of its correlation
functions.
 The phase space version of the CTP generating functional leads to the
implementation of a Wigner-Weyl transforms, that gives a
description of quantum theory solely in terms of phase space
histories. These results allow the identification of an algorithm
for going  to the  classical (stochastic) limit for a generic
quantum system.

\end{abstract}

\renewcommand {\thesection}{\Roman{section}}
 \renewcommand {\theequation}{\thesection. \arabic{equation}}
\let \ssection = \section
\renewcommand{\section}{\setcounter{equation}{0} \ssection}
\pagebreak

\section{Introduction}

\subsection{ Canonical vs covariant}
Physical systems can be described in two different ways, depending on one's  attitude
towards  time evolution. The first description can be called ``canonical'': it focuses on
properties of a system at a single
moment of time and studies how these  properties change.
 It, therefore, provides an evolutionary
 picture of physical phenomena.
The other type is best described as ``covariant'': its main objects are histories of the
 physical system. Its main   aim  is to find  criteria that determine which of them are
 realizable. As such, this description   provides a timeless and (in a sense)
 teleological picture of physical processes.

In classical mechanics the ``canonical'' description is Hamilton's formalism.
States of the system correspond to points of
  the phase space, which is a symplectic manifold.
  Time evolution is implemented by the
 action of an one-parameter group of  symplectic transformations.
Alternatively,
 one can start from the action principle, which  provides the covariant
 description of classical mechanics. Histories are paths, and
the physically realized are the ones that minimize the action
subject to fixed boundary conditions .

These two approaches  also appear in classical probability theory.  A physical system at a
moment of
 time   is described by  a probability distribution on a  space $\Omega$ of
elementary alternatives. We then study how this
 distribution evolves in time: the evolution law  is  a linear partial differential equation,
like the Fokker-Planck equation.
The ``covariant'' description of probability theory is provided by the theory of stochastic
 processes. Here, histories are paths on $\Omega$ and the physical information is encoded
in a probability measure $d \mu$  in the space of all histories;
it incorporates information about both initial conditions and
dynamics.

Quantum theory  was developed in the ``canonical'' framework. The probabilistic information
about  a system is encoded in a Hilbert space vector, or more generally a density matrix.
Its time
 evolution is given by an one-parameter group of unitary transformations: this is
 equivalent to Schr\"odinger's equation. The general structure is very similar to classical
 probability theory, except for the fact that the  observables do  not form a commutative
 algebra.

\subsection{Quantum mechanical histories}

When one tries to construct a "covariant" description of quantum theory, a problem immediately
 arises:  {\em the  natural probability measure for histories is not additive}. This is due to
 the fact, that quantum theory is based on amplitudes. When one constructs probabilities out
 these amplitudes, interference between histories appears.

In general, a   history  corresponds to properties of the physical
system at successive instants of time. Since in quantum theory a property (or a proposition
 about it) is represented by a projection operator, a discrete-time history $\alpha$ will
correspond to a string $\hat{\alpha}_{t_1}, \hat{\alpha}_{t_2},
\ldots \hat{\alpha}_{t_n}$ of projectors, each labelled by  an
instant  of time. From them, one can construct the class operator
\begin{equation}
\hat{C}_{\alpha} = \hat{U}^{\dagger}(t_1) \hat{\alpha}_{t_1} \hat{U}(t_1)
\ldots \hat{U}^{\dagger}(t_n) \hat{\alpha}_{t_n} \hat{U}(t_n)
\end{equation}
where $\hat{U}(s) = e^{-i\hat{H}s}$ is the time-evolution
operator. The probability for the realization of this history is
\begin{equation}
p(\alpha) = Tr \left( \hat{C}_{\alpha}^{\dagger}\hat{\rho}_0  \hat{C}_{\alpha} \right),
\end{equation}
 where $\hat{\rho}_0$ is the density matrix describing the system at time $t = 0 $.

But this expression does not define a probability measure in the space of all histories,
because  the Kolmogorov additivity condition cannot
 be satisfied: if $\alpha$ and $\beta$ are exclusive histories
 and $\alpha \vee \beta$ denotes their conjunction as propositions, then it is not true that
\begin{equation}
p(\alpha \vee \beta ) = p(\alpha) + p(\beta) .
\end{equation}
 The
histories formulation of quantum theory does not, therefore,  enjoy the status of a genuine
 probability theory.

\subsubsection{The consistent histories interpretation}
The  formalism sketched above was developed as a part of the
consistent histories approach to quantum theory, by Griffiths,
 Omn\'es, Gell-Mann and Hartle \cite{Gri84, Omn8894, GeHa9093, Har93a}. In this approach,
  the problem of the
non-additivity of the probability measure  is addressed by the remark that an
 additive probability measure {\it is} definable, when we restrict to
particular  sets of histories.
 These are called {\it consistent sets}. They are more conveniently
defined through the introduction of a new object: the decoherence
functional. This is a complex-valued function of a pair of
histories given by
\begin{equation}
d(\alpha, \beta) = Tr \left( \hat{C}_{\alpha}^{\dagger} \hat{\rho}_0 \hat{C}_{\beta} \right).
\end{equation}
A set of exclusive and exhaustive alternatives is called
consistent, if for all  pairs of different histories $\alpha$ and
$\beta$ in the set, we have
\begin{equation}
d(\alpha, \beta) = 0 .
\end{equation}
In that case one can use equation (1.2) to assign a probability measure to this set. The
consistent histories interpretation then proceeds by postulating that any prediction or
 retrodiction, we can make based on probabilities
{\it has always to
 make  reference to a given consistent set}. This  leads to  counter-intuitive
   and arguably unphysical
situations of getting mutually incompatible predictions, when
reasoning within different consistent sets \cite{DoKe96, Kent97}.
 The predictions of this theory are therefore contextual: but in
 any case, this  is a general feature of  all realist interpretations of quantum theory.

Even if  the formalism of quantum mechanical histories was originally introduced as part of the
consistent histories approach, it is conceptually distinct.
The same formalism can be viewed in the light
 of any other interpretational scheme.
The Copenhagen interpretation, for instance,  would view the
non-additivity of the probability measure  in a neutral light. The
expression (1.2) describes the statistics of an ensemble of
time-ordered sequences of measurements. There would be  no a
priori theoretical reason for the statistics to correspond to a
genuine probability measure.

In this paper, we shall focus on the formal aspect  of quantum mechanical histories. We do not find necessary to
 commit to any particular interpretation: we only assume that all physical information about
 probabilities and interference of histories is encoded in the decoherence functional,
something very explicitly shown by Gell-Mann and Hartle. It is not
our aim to insist on how this information can be extracted: both
the logic of consistent sets  and the Copenhagen stance provide
ways of doing this. Perhaps these ways do not exhaust the
 physical content of the theory --we have argued this in reference \cite{Ana00f}, but each of
 them is separately adequate to account
  for all minimal predictions of standard quantum theory.

We    view the histories formalism  simply as the covariant version of quantum
theory.
 As such, it  incorporates features of  the covariant formulation of  both classical mechanics
 and probability theory. But interference of probabilities highlights its quantum nature, and
 for this reason we shall pay particular attention to the structure of the
decoherence functional.

\subsubsection{Temporal logic histories}

We shall work in the context of temporal logic histories. This is a scheme  initiated by
 Isham \cite{I94, IL94}: its main point is that the quantum logic  is preserved in the histories theory if we
represent a history proposition $(\alpha_{t_1}, \ldots, \alpha_{t_n})$  by a projection
operator  on  a tensor product of the Hilbert spaces of the canonical theory
 ${\cal V} = \otimes_{i} H_{t_i}$. This history proposition will then be written as
 $ \alpha  = \alpha_{t_1} \otimes \ldots \otimes \alpha_{t_n}$. This construction is completely analogous
to the construction of the space of classical histories as a {\it Cartesian product} of
single-time sample spaces.

In this formulation a self-adjoint operator on ${\cal V}$
represents a history observable for the physical system. As in
 any covariant theory, more general observables  can be defined. They correspond to
  time-averages and  include, for instance, an {\it action operator}.

One of the great strengths  of this formalism  is found in its temporal
structure. It was shown by Savvidou \cite{Sav99a, Sav99b},  that one can mathematically
 distinguish between two
 qualities of time: its partial ordering properties ( the notion
of before and after) and its status as a dynamical parameter  in the equations
of motion.

To see this, examine equation (1.1) for the class operator $C_{\alpha}$ entering the
 expression for the decoherence functional. There, time appears in two places: as an index
of the projectors $\hat{\alpha}_t$ and as the argument of the unitary operators $\hat{U}(t)$.
In its former status, it is
 purely a kinematical parameter labelling the moment  upon which a proposition is asserted.
 Its function is to determine the order upon which propositions are asserted, in the sense that if $t_1 \leq t_2$
the projection operator $\hat{\alpha}_{t_1}$ will appear on the left of the operator $\hat{\alpha}_{t_2}$ in the
equation (1.1) for $\hat{C}_{\alpha}$.
  In  its latter status as the argument of the unitary operators, time  is
the  parameter of the Heisenberg-type  evolution. It, thereby,
implements the dynamics of the system.

These two roles of the time
 parameter are completely disentangled, when we view histories in the tensor product Hilbert
 space ${\cal V} = \otimes_t H_t$. This is an intriguing property, since it allows us to
 {\em mathematically distinguish} between the two conceptually distinct roles by which time
  appears in physical theories: in the form of a causal structure, that determines the
order of events and in the form of the parameter by which change is
 effected in a physical system.

Indeed in the histories formalism  there appear  two mathematically distinct
 laws of time transformation.
 The partial ordering aspect of
 time is manifested in translations of the form $H_t \rightarrow H_{t+a}$, by which
a property asserted at time $t$ is translated to the same property
at time $t+a$.
 At  the continuum limit these transformations  are
generated by the kinematical part of an action
 operator. Dynamical time transformations  are equivalent
to a separate unitary transformation for  each single-time
Hilbert space $H_t$. They correspond to the Hamiltonian part of the action.

This distinguishing presence of two laws of time transformation is an important physical
 principle, that
 will provide an   guideline for the construction of history theories, in the case where
  the canonical formalism does not provide sufficient insight. In retrospect, one can
   see that this distinction is present in {\em all} physical theories that are
   formulated in a covariant (histories) fashion \cite{Sav99b}.

\subsection{This paper}

Since our aim is to show how histories provide a covariant
formulation of quantum theory, we need to go beyond the
discrete-time description that is usually effected: time, in
physics, is a continuum.  Continuous-time histories have  been
introduced in \cite{IL95} and further studied
 in \cite{ILSS98,Sav99a,Ana00,SavAn00}. This work relied on the use of a Fock
 space for the history Hilbert space,
 which is only justified if  the Hamiltonian is  quadratic.

The first aim of this paper is, therefore,  to explore the nature of continuous time in this
 framework. In particular, we highlight the structures that arise
in the probability assignment. The analogy with stochastic processes is quite helpful in
this regard, both at a conceptual and at a technical level.

In section 2 we explain how a continuous-time Hilbert space with
physically interesting observables can be constructed. We then
  analyze the decoherence functional: we show that it can be
  decomposed in  a way that respects the two laws of time transformation.
In fact, its components reflect
 the distinction between the geometric and the dynamical phase of canonical
  quantum theory \cite{AnSav00}.
Finally, we discuss the time-reversal transformations,
which are substantially different from the ones of  standard quantum theory.

In section 3 we study the phase space structure of histories. This is incorporated in the
 quantum theory through the use of the history group, the history  analogue of the
 canonical group. The history Hilbert space carries one of its representations. This allows the
   identification of self-adjoint operators in this Hilbert space with objects
 that have a classical phase space analogue. We explain, how one can construct such
 representations from the knowledge of the canonical theory.

The analogy with classical probability suggests  that one should treat  the decoherence
 functional as the quantum analogue of a classical probability measure. In this sense its
 ``Fourier transform''  yields the analogue of the generating function of classical
 probability: this is the closed-time-path (CTP) generating functional, first introduced by
 Schwinger \cite{Schw61}. We show how to construct this object for phase space histories.
  This construction
suggests that  the Wigner transform is of relevance: it enables us to write the decoherence
 functional as a complex-valued  measure on the space of phase space paths and provides a
 picture of quantum theory that makes reference  only to classical
objects. One of the merits of this construction is that it provides an algorithmic procedure for passing into the
 classical limit of  generic quantum theories.

In the final section we  review our results. We argue that the formalism is flexible enough to accommodate a large number
of applications in different fields. In particular, we stress the importance of our results
as part of the developing continuous-time histories programme.

 Overall, our attitude  is to highlight
 similarities of structures between the histories formalism and more familiar physical
formalisms, such as stochastic processes or canonical quantum theory.

\subsubsection{Notation}
In the following, our expressions will make reference to two
different types of Hilbert space: canonical ones and history ones.
We adopt the following conventions: we will use the braket
notation to denote vectors of both types of Hilbert space. But  we
will insert a subscript in the ket denoting a {\em canonical}
Hilbert space. Hence, for instance, $| \psi_t \rangle_{H_t}$ will
denote a vector on the canonical Hilbert space $H_t$, while $|\psi
\rangle$ will denote a vector on a history Hilbert space ${\cal
V}$.

Also, operators on canonical Hilbert spaces will carry a hat, while the history ones
will be unhatted.

As already seen in the introduction, we use the same symbol (small
Greek letters) to denote both a proposition and the projector that
represents it. We let the meaning be determined by the context.

The notation in section 3 is more complicated, because of the many
spaces involved. Points of the (linear) phase space $\Gamma$ will
be denoted as $(q,p)$. But there is also the space
$\tilde{\Gamma}$, which is the vector space dual of $\Gamma$ and
(if $\Gamma$ is a Hilbert space) isomorphic to it. Points on this
space will be denoted as $(\chi, \xi)$: they correspond to
elements of the canonical group or labels of coherent state
vectors. The latter will often be denoted as $| z \rangle$. The
inner product in these spaces, will be denoted by a dot: we will
write invariably
 $q \cdot p$ or $q \cdot \xi$, without reference to whether
 the arguments are elements of  $\Gamma$ or $\tilde{\Gamma}$. In fact, we shall
mostly ignore their distinction.

Paths on $\Gamma$ will be denoted as $(q,p)(\cdot)$, or $ t
\rightarrow (q_t,p_t)$, or  simply $\gamma$. Paths on
$\tilde{\Gamma}$, corresponding to coherent state histories by
$(\chi, \xi)(\cdot)$, or $t \rightarrow (\chi_t,\xi_t)$, or simply
$z(\cdot)$. We will write $(q,\xi) = \int d \mu(t) q_t \xi(t)$.
When we want to emphasize that $\xi$ also acts as a smearing
function on $q_t$ we will denote the same object as $q_{\xi}$.

\section{Continuous-time histories}
\subsection{The basic structure}

The temporal logic histories scheme is based on ideas from quantum logic. It seeks
 to represent
  the set of all history propositions about a physical system with elements of a lattice, that
contains the information about the temporal structure \cite{I94}.

Let us denote by $T$ the set of  all instants of time (this can be
either discrete, or the real line ${\bf R}$ or a subset of ${\bf
R}$). Standard quantum theory is recovered, when we consider that
history propositions correspond to projection operators on a
Hilbert space ${\cal V}$, given by the tensor product $\otimes_{t
\in T} H_t$. Here $H_t$ is
 a {\em copy} of the Hilbert space of the canonical theory indexed by $t$.

Self- adjoint operators on this Hilbert space correspond to history observables.

As an example,
 let us consider the case where
 $T$ is a finite set.  Let  $\hat{A}$ be a bounded operator on the  Hilbert space $H$ of
 the canonical theory, and let us denote by  $\hat{A}_t$  its copy on
a Hilbert space $H_t$.  Then
we can define the {\it product operator} $\otimes_{t \in T} \hat{A}_t$
on ${\cal V}$.

If $\hat{A}_t$ is unit everywhere, but a single point $t \in T$, then we shall denote the product operator on ${\cal V}$ as $A_t$.
If $f:T \rightarrow {\bf R}$ then we can define the {\it time-averaged} operator $A_f$ as
\begin{equation}
A_f = \sum_{t \in T} f(t) A_t
\end{equation}
It corresponds to the average in time of the family $t \rightarrow \hat{A}_t$, with a weight given by the function $f$.
We can easily verify the following identity. If $\hat{A}$ is a self-adjoint operator on $H$, then its time-averaged counterpart
on ${\cal V}$ satisfies
\begin{equation}
e^{iA_fs} = \otimes_{t \in T} e^{i \hat{A}f(t) s}
\end{equation}
We shall use this identity to {\em define} time-averaged operators
in the continuous-time case.

Note also that for the case of projection operators, a map $\hat{\alpha} \rightarrow \alpha_t$ provides
a continuous embedding of the lattice of propositions at a single moment of time to the lattice of history
propositions.

The probabilistic content of  the theory is contained in the decoherence functional.
 This is  assumed to satisfy the following conditions
\begin{eqnarray}
d(1,1) &=& 1 \nonumber \\
d(\alpha, \beta) &=& d^*(\beta, \alpha) \nonumber \\
d(0, \alpha) &=& 0 \nonumber \\
d(\alpha + \beta, \alpha') &=& d(\alpha, \alpha') + d(\beta, \alpha') \nonumber \\
d(\alpha, \alpha) &\geq& 0
\end{eqnarray}
In general, there exists a  class of operators $X$  on ${\cal V} \otimes {\cal V}$,
such that a decoherence functional can be written as \cite{ILS94,RuWr97}
\begin{equation}
d(\alpha, \beta) = Tr_{{\cal V} \otimes {\cal V} } \left( X \alpha \otimes \beta \right)
\end{equation}

When the space $T$ is finite, the construction of the tensor product Hilbert space is
 straightforward and equation (1.4) can be used to construct the  decoherence functional.
  The question arises then, how  one deals with  continuous time. This is the case  when $T$
is a closed subset of the real line. For particular systems the construction of
 such Hilbert spaces has been carried in \cite{IL95}. For more general cases, we believe it
 is instructive to look at the analogous situation in the classical setting.

\subsection{Classical stochastic processes}
Let us assume we have a classical system that at a moment of time is described by a sample
 space $\Omega$. Let us also consider the space $T$ of time instants to be a closed subset
 of the real line , say $[0,a]$. The space of histories $\Pi$ is then some suitable subset
 of the set $\Omega^T$  of all measurable maps $\gamma: T \rightarrow \Omega$. If $\Omega$
 is a vector space  one can  define a norm on $\Omega^T$, and take
as $\Pi$ the subspace of $\Omega^T$, that contains  paths with finite norm.

A function $f$ on $\Omega$ defines a family of functions $F_t$ on $\Pi$ by
\begin{equation}
F_t(\gamma) = f(\gamma(t))
\end{equation}

As a stochastic process, we usually define a triplet consisting of the space $\Pi$,
 a family $F_t$ and a  measure $d \mu$ on $\Pi$.  The issue is how to  construct
physically interesting measures on $\Pi$, which is an infinite dimensional
function space.

This is effected as follows:
Let $d x$ be for brevity a
natural integration measure on $\Omega$ (say a Lebesque measure). Let $T = [t_0, t_f] $ be an
 interval and let  us also consider a discretization $ I = \{t_0, t_1, \ldots
 t_n = t_f \} $ of  $T$. Then define the space of discrete time histories $\Omega^I =
\times_{t_j \in I} \Omega_{t_j} $, which is a finite dimensional manifold. This admits the
 measure $\prod_{t_j \in I} dx_{t_j}$. Any probability distribution
$p_I(x_{t_1}, \ldots, x_{t_n})$ on $\Omega^I$ defines a measure $
d \mu (x) = p_I(x_{t_1}, \ldots x_{t_n})
 \prod_{t_j \in I} d x_{t_j}$.

As we  consider   all possible discretizations $I$ of $T$, we can
 encode a choice of probability measure for each discretization in a hierarchy of positive
 functions
\begin{eqnarray}
&{}&p_1(x,t) \nonumber \\
&{}&p_2(x_1,t_1;x_2,t_2) \nonumber \\
&{}&.... \nonumber \\
&{}&p_n(x_1,t_1; \ldots ; x_n,t_n) \nonumber \\
&{}&....
\end{eqnarray}
These have to be symmetric with respect to interchange of their $(x,t)$ arguments.

Now, the fundamental theorem of Kolmogorov asserts the following: If a hierarchy  of functions
 as above, satisfies the {\it additivity condition}:
\begin{equation}
 \int dx_n  p_n(x_1,t_1; \ldots ;x_{n-1}, t_{n-1}; x_n,t_n)  =
p_{n-1}(x_1,t_1; \ldots  ; x_{n-1},t_{n-1})
\end{equation}
then there exists an essentially  unique probability measure $d \mu ( \cdot )$ on $\Omega^T$ such that it gives
 the correct discrete time probability measures, i.e. for each partition $I$ ,
$j_I^* d \mu = d \mu_I$, where $j_I$ is the natural injection map $j_I : I \rightarrow T$.

Kolmogorov's proof is standard textbook material and is one
instance of a general categorical construction of taking the
inductive limit. The essential  point   in the proof is the fact
that $j_I$ is a measurable map (with respect to the Borel sets of
$T$)  and as such it respects the measurable structure in the
definition of $d \mu$.

{\it Hence a probability measure is defined for continuous time,
while making reference only to discrete time expressions}.  This
is the theorem that we will try to employ, in order to
 construct the decoherence functional for continuous- time histories.

\subsection{ The continuum limit}

\subsubsection{The Hilbert space}
The first objective would be to define a suitable version of the Hilbert space
 ${\cal V} = \otimes_{t \in T} H_t$.  This expression cannot be taken literally, for a
continuous tensor product of Hilbert spaces leads to a
non-separable Hilbert space. What we will do is a generalization
of an idea that has been applied to "continuous tensor products"
of Fock spaces \cite{IL95}.

 Consider the space $B(T,H)$ of continuous maps $|\psi( \cdot ) \rangle$ from $T$ to $H$.
In fact, we can start our construction considering only measurable maps. But since
we will later want to define Stieljes integrals, we should impose the restriction that the  maps are  of
{\em bounded variation}, i.e. they  satisfy the following property:
  \\ \\
For any finite discretization of $T$: $\{ t_0 \leq  t_1, \leq
\ldots \leq  t_i \leq \ldots \leq  t_n \}$, the sum $ \sum_{i=1}^n
|| \psi_{t_i} - \psi_{t_{i-1}}||_H$ is finite.
\\ \\
 Assume that $T$ has
 a measure $d \mu(t)$, which in the standard case should be taken as $ \frac{d t}{\tau}$.
 Here $\tau $ is a  time parameter that makes the measure dimensionless.
If $T$ is compact it can be used to normalize the measure $\mu(T)
= 1$.

Then define the inner product
\begin{equation}
\langle \psi( \cdot )| \phi( \cdot ) \rangle = \prod_{d \mu(t)}  \langle \psi_t| \phi_t \rangle
:= \exp \left( \int d \mu(t) \log [  \langle \psi_t| \phi_t \rangle_{H_t} ] \right),
\end{equation}
where it is understood that the inner product vanishes if $ \langle \psi_t| \phi_t \rangle_{H_t} = 0$
in a subset of $T$ that is not of measure zero, and that the
logarithm takes values on the principal branch.

This space has then a norm $|| \psi( \cdot )|| = (\langle \psi(
\cdot )| \psi( \cdot ) \rangle)^{1/2}$. We identify two elements
$\psi_1( \cdot ), \psi_2( \cdot )$ of $B(T,H)$, if $||\psi_1(
\cdot ) - \psi_2 ( \cdot )|| = 0$. {\em This identification makes
the resulting Hilbert space separable}.

 Let us, suggestively,  denote the vector space  we obtained after identification,
 as $\times_{t \in T} H_t$.
To construct $\otimes_{t \in T} H_t$ we  consider the space of all formal
linear combinations $\sum_i c_i | \psi_i( \cdot ) \rangle$. Here  $i$ runs over a finite set, $c_i \in {\bf C}$,
and $ \{| \psi_i ( \cdot ) \rangle \}$ is  a finite set of vectors of $\times_{t \in T}  H_t$.
 On the space of these formal linear combinations we define the inner product as
\begin{equation}
\sum_i \bar{c'}_i c_i \langle \psi'_i ( \cdot )| \psi_i( \cdot ) \rangle
\end{equation}
and close this space with respect to the norm. We have thus  defined a Hilbert space
$\otimes_{t \in T} H_t$. Note
that the time parameter $\tau$ enters explicitly into the definition.

The vectors $| \psi( \cdot ) \rangle$ form a total set of $\otimes_{t \in T}H_t$.
 As such, we can define  operators on the history Hilbert space by their
 action on these vectors.

Some properties of this construction are easy to see. For instance
\begin{equation}
\otimes_t e^{\lambda_t} | \psi_t \rangle_{H_t} = e^{\int d \mu(t) \lambda(t)} | \psi( \cdot ) \rangle .
\end{equation}

Also, if $T_1$ and $T_2$ are two disjoint
  subsets of ${\bf R}$ with non-zero measure, then
\begin{equation}
\otimes _{t \in T_1 \cup T_2} H_t = (\otimes_{t \in T_1}H_{t}) \otimes (\otimes_{t \in T_2} H_t)  .
\end{equation}

\subsubsection{Time-averaged observables}
Let $\hat{A}_t$ is a continuous family of bounded    operators on $H$ indexed by $t$.
 Then one can define the product operator $\otimes_{t \in T} \hat{A}_t$
  by its action on $|\psi( \cdot ) \rangle$
\begin{equation}
\left(\otimes_{t \in T} \hat{A}_t\right) | \psi( \cdot ) \rangle =
\otimes_{t \in T} \left(\hat{A}_t |\psi_t \rangle_{H_t}\right).
\end{equation}
This definition is extended by linearity to the whole Hilbert
space. However, one has to restrict  the families $t \rightarrow
\hat{A}_t$. We have
\begin{eqnarray}
||  \left( \otimes_{t \in T} \hat{A}_t \right)  | \psi( \cdot ) \rangle ||^2 =
\exp \left( \int d \mu(t) \log (  \langle  \psi_t |\hat{A}_t^2 | \psi_t \rangle_{H_t} \right) \nonumber \\
\leq \exp \left( \int d \mu(t) \log ( ||A_t||^2 \langle \psi_t| \psi_t \rangle_{H_t})
\right) =
 e^{\int d \mu(t) \log( ||\hat{A}_t||^2  )} \langle \psi( \cdot )| \psi( \cdot ) \rangle ,
 \end{eqnarray}
hence one has to demand that $\int d \mu(t) \log(||\hat{A}_t||^2 ) \leq \infty$.
If $T$ is  a compact subset of ${\bf R}$,  this holds automatically provided the map $t \rightarrow ||\hat{A}_t||$ is measurable.
 If $T$ is non-compact, e.g. the whole of ${\bf R}$,  the right hand side is not finite and
one has to additionally demand that $\hat{A}_t = 1 $ outside some compact subset of ${\bf R}$, or that $||\hat{A}_t - \hat{1}||_H$ falls to zero
sufficiently rapidly.

It is easy to see that
\begin{equation}
Tr_{{\cal V}}\left(\otimes_{t \in T} \hat{A}_t \right)= \prod_{t \in T} \left( Tr_{H_t}\hat{A}_t \right) = \exp ( \int d \mu(t) \log Tr_{H_t}\hat{A}_t)  .
\end{equation}

Having defined the product operators we can define time-averaged
observables, by exploiting equation (2.2). Let $\hat{A}$ be a
bounded self-adjoint operator on $H$. We can write  the family  $t
\rightarrow \hat{U}_t(s) = e^{i\hat{A}f(t) s}$ of unitary
 operators and construct the product operator
$U_f(s) =\otimes_{t \in T } \hat{U}_t(s)$. This is well-defined if
$f(t) \neq 0$ only within a compact subset of ${\bf R}$ and
corresponds to an one-parameter group of unitary operators on
${\cal V}$. By Stone's theorem, if the matrix elements of this
operator  are continuous functions of $s$ at $s = 0 $,  there
exists a self-adjoint operator $A_f$ such that $U_f(s) =
e^{iA_fs}$.

It is easy to check,  that
\begin{equation}
\langle \phi( \cdot )|U_f(s)| \psi( \cdot ) \rangle = \exp \left( \int d \mu(t)
 \log ( \langle \phi_t|e^{i\hat{A}f(t)s}| \psi_t \rangle) \right)
\end{equation}
is a continuous function of $s$ at $s = 0$, when the operator $\hat{A}$ is bounded.
Thus, given suitable functions $f$,  a self-adjoint
operator representing the time average of $\hat{A}$
is well defined on ${\cal V}$.

\subsubsection{Unbounded operators}
The construction of time-averaged counterparts to unbounded operators on $H$ is more complicated. From equation (2.15)
 we see that even if the matrix elements
$\langle \psi_t|e^{i\hat{A}f(t)s} | \phi_t \rangle_{H_t}$ are continuous functions of $s$, there is no guarantee that so will be the integral.

Also if  $\hat{A}$ is unbounded, there exist vectors $|\psi_t
\rangle$, for which the action of $\hat{A}_t$ is not defined,
hence one cannot write $ |\langle \phi_t|e^{i\hat{A}f(t)s}| \psi_t
\rangle| \leq c |s| $, which would be sufficient to prove
continuity. There is {\em no guarantee} that the time-average of
an unbounded operator is definable.

 This is unfortunate, because
in physical situations  we are interested in operators like position, or momentum, or the
 Hamiltonian, that are typically unbounded. This failure is due to the fact that the Hilbert
 space $\otimes_{t \in T}H_t$ is still very large. In concrete physical situations one
 should identify the histories Hilbert space ${\cal V}$ with a closed linear subspace
 of $  \otimes_{t \in T}H_t$.

One has to choose this closed linear subspace, in such a way that the tensor product structure
 is preserved. The simplest way is to restrict the set of vectors  that can be used
 to construct the "paths"
$| \psi( \cdot ) \rangle$ to a subset ${\cal L}$ of $H$. This set ${\cal L}$ has to be sufficiently
large to be able to capture all physical information from $H$ (it cannot be a subspace of $H$),
but small enough to allow interesting operators to be definable on the history Hilbert
space. A good choice for ${\cal L}$ is an overcomplete and continuous family of vectors, like the
coherent states.

Having chosen ${\cal L}$, the construction proceeds as before, only we substitute $B(T,H)$
with
 $B(T,{\cal L})$:  the
space of all continuous maps from $T$ to ${\cal L}$.
  It is easy to check that the resulting Hilbert space is a closed linear subspace
 of $\otimes_{t \in T} H_t$.

If we demand that a particular unbounded
 operator $A$ exists (time-averaged) in our Hilbert
space, it would be necessary to take ${\cal L}$ consisting of vectors in the
domain of $A$. In that case the matrix elements (2.15)
 would be a continuous function of $s$
 and (by Stone's theorem) $A_f$ would exist. We shall see how  this construction works in more detail, in section 3.
  In this section,
we shall work with  the larger Hilbert
space $\otimes_{t \in T} H_t$. All   results we obtain will be  valid for any of its physically relevant subspaces.

\subsubsection{The decoherence functional}
If $T$ is compact, one can choose $A_t = A$ for all $t$ and therefore interpret
$A_f$ as the time average  of the quantity associated to $A$. But, if we try to
define an operator on ${\cal V}$, that corresponds to an observable at a {\it sharp} moment
of time, we run into problems. Since a point in the real axis is of measure zero,
 an observable defined at a sharp moment of time can exist only if we can take $f$ to be
 a delta function. This is unacceptable in our construction.  We conclude  that one {\em cannot embed
continuously the lattice of single-time propositions into the lattice of history propositions}, in the
case of continuous time.

Let us now examine the possibility of defining a decoherence functional for continuous-time
histories as a continuous limit of the discrete-time expression (1.4).
Let us assume a partition $I = \{ t_1, \ldots, t_n \}$  of an interval $T$ of the real line
 and a proposition $ \alpha =  \alpha_{t_1} \otimes \ldots \otimes \alpha_{t_n}$ that is a projector
 operator on $H^I = \otimes_{t_i \in I} H_{t_i}$. Then one can construct the class operator
 $\hat{C}_{\alpha}$ defined on one one copy of $H$ as in equation (1.1).
The value of decoherence functional $d_{I,I'}$  between a history on $H^I$ and another on some other discrete-time
Hilbert space $H^{I'}$ is given by equation (1.4).

The aim is to generalise Kolmogorov's theorem in this histories setting.
We want to construct a bilinear, hermitian, additive map on the space
$P({\cal V}) \times P({\cal V})$ (by $P(H)$ we mean the lattice of projectors on the Hilbert space $H$). If we
consider then a pair of discretisations $I$ and $I'$ of $T$, we can costruct the Hilbert
spaces $H^I$ and $H^{I'}$. The point is whether there exist an injection map
$ j_{I,I'} : H^I \times H^{I'} \rightarrow {\cal V} \times {\cal V}$; if this exists and preserves the
lattice structures
then  Kolmogorov's  proof goes
through and the decoherence functional $d$ on $H^T$ exists
as an inductive limit of the decoherence functional defined on $H^I \times H^{I'}$ for all choices of $I$ and $I'$.
We would also have  $ d_{I,I'} = j_{I,I'}^*d_T$.

For the map to be lattice-preserving it would have to be continuous. But, we showed
earlier, that this cannot be true for a single moment of time. The map $j_{I,I'}$  might
 be continuous in the weak
 topology, but this is insufficient to define an order preserving
 map. Recall that the continuity
 of the Hilbert space enters in a decisive point in the definition of the lattice of
 propositions: a projection operator corresponds to a {\it closed} linear subspace.
Hence Kolmogorov's theorem does not go through in this case.

But  if we restrict to an Abelian
 sublattice, (for instance, to  propositions   about position) the map
 $j_{I,I'}$ does not need to be a continuous, linear map, but simply a measurable map from
  the spectra of the corresponding operators
 ${\bf R}^I \times {\bf R}^{I'} $ to ${\bf R}^T \times {\bf R}^T$.
This clearly exists;
 it is the same as in the  case of classical probability theory.

 We therefore conclude  that {\em one cannot write the decoherence functional
 for continuous time, as a limit of discrete-time ones}, unless one restricts to Abelian
 subalgebras. We might have a continuous- time decoherence functional for each subalgebra,
 but not one defined on the whole of $P({\cal V})$. We shall return to this issue again and
 propose two different ways, by which the decoherence functional can be defined.

\subsection{The structure of the decoherence functional}
The presence of two laws of time transformation is an important structural feature of history
theories. In this section, we shall show how they are manifested in the probability assignment.
We shall see that the decoherence functional for discrete time can be written in such a way, that
these two notions of time are distinctly represented. This is a feature, that in the canonical
theory is reflected in the distinction between geometric and dynamical phase \cite{AnSav00}. And this feature
we shall attempt
to generalise in the continuous-time case.

For simplicity we shall consider a special class of decoherence functionals. They are of
 the type
(1.4), but with $\hat{\rho}_0$ corresponding to  a pure  state. This means that we can absorb the projector into the
 initial
state as part of the definition of each history and as such write the decoherence functional
 in the form
\begin{equation}
d(\alpha, \beta) = Tr_H (\hat{C}^{\dagger}_{\alpha} \hat{C}_{\beta} ).
\end{equation}
Clearly one of the single-time projectors has to be trace-class if the above expression
is to be finite.
Equation (2.16)  can be written as \cite{ILS94}
 \begin{equation}
d(\alpha, \beta) = Tr_{H \otimes H} ( Z \hat{C}^{\dagger}_{\alpha} \otimes \hat{C}_{\beta}),
\end{equation}
where $Z$ is an operator on $H \otimes H$ given by
\begin{equation}
Z(|i\rangle \otimes |j\rangle) = | j \rangle \otimes | i \rangle .
\end{equation}

One can write $Z = \sum_{rs} \hat{A}^{ rs} \otimes \hat{A}^{\dagger rs}$ in terms of a basis on H, where
$\hat{A}^{rs} $ is an operator on $H$ with matrix elements
\begin{equation}
\langle k|\hat{A}^{rs}| i \rangle = \delta_{ks} \delta_{ri} .
\end{equation}

Let us now assume that both histories are defined in the same  instants of time
 $t_0, t_1, \ldots
t_n$. Let us for simplicity take $t_0 = 0$. The corresponding history Hilbert space is then ${\cal V} = \otimes_i H_{t_i}$.

Let us also
write the {\em boundary Hilbert space} $\partial {\cal V} = H_{t_0} \otimes H_{t_n}$. The indices $rs$
of the operators $\hat{A}^{rs}$ are then indices corresponding to $\partial {\cal V}$.
It is easy to verify that the expression (2.17) can be written as a trace over the boundary
Hilbert space \cite{Ana00}
\begin{equation}
d(\alpha, \beta) = Tr_{\partial {\cal V}} \left( c(\alpha) c^{\dagger}(\beta)
\right),
\end{equation}
where $c(\alpha)$ is an operator on $\partial {\cal V}$ defined by
\begin{equation}
c(\alpha) = Tr_H \hat{A} \hat{C}_{\alpha}^{\dagger} .
\end{equation}
$\hat{A}$ denotes here a map from $H$ to ${\cal V}$. It is easy now to write $c(\alpha)$ as a
trace over the history Hilbert space, through the introduction of the {\it unitary} operator ${\cal S}$ on ${\cal V}$
\begin{eqnarray}
{\cal S} |v_{t_0} \rangle | v_{t_2} \rangle \ldots | v_{t_n} \rangle =
|v_{t_n} \rangle | v_{t_0} \rangle \ldots | v_{t_{n-1}} \rangle .
\end{eqnarray}
Indeed, since $\alpha = \hat{\alpha}_{t_0} \otimes \hat{\alpha}_{t_1} \otimes \ldots \otimes \hat{\alpha}_{t_n}$,   we can write
\begin{equation}
 c(\alpha) = Tr_{{\cal V}} \left( {\cal AS} {\cal U}^{\dagger} \alpha  {\cal U} \right) ,
\end{equation}
where
\begin{eqnarray}
{\cal U} = \hat{U}(t_0) \otimes \hat{U}(t_1)  \ldots \otimes \hat{U}(t_n), \\
{\cal A} = \hat{A} \otimes 1 \otimes \ldots \otimes 1 .
\end{eqnarray}
This accomplishes the task of writing the decoherence functional in such a way as the two different
notions of time are made manifest. The operator ${\cal U}$ clearly contains the dynamics. The operator
${\cal S}$ induces a transformation that takes from a single-time Hilbert space to the next one.
Finally, the operators $A$  incorporate the information about the beginning and the end of the
interval. Had we kept the initial density matrix, $\hat{A}$ would explicitly depend upon it. In that case the
analogue of equation (2.19) would be
\begin{equation}
\langle k |\hat{A}^{rs}|i \rangle =  \delta_{ks}
(\rho_0^{1/2})_{ri}.
\end{equation}

 \subsection{ The continuum limit}
Let us now examine whether one can construct these operators in the continuous-time Hilbert
space $\otimes_{t \in T} H_t$, which we defined earlier.

The operator ${\cal U}$ is relatively easy to define. It would act on a vector $| \psi ( \cdot ) \rangle$
as
\begin{equation}
{\cal U} | \psi ( \cdot ) \rangle = \otimes_{t \in T} \left(e^{-i\hat{H}t} | \psi_t \rangle_{H_t}\right).
\end{equation}
This would have as matrix elements
\begin{equation}
\langle \phi( \cdot ) | {\cal U} | \psi( \cdot ) \rangle = \exp
\left( \int d \mu(t) \log \langle \phi_t| e^{-i\hat{H}t}| \psi_t
\rangle \right).
\end{equation}
According to our previous analysis this  is a genuine unitary
operator on ${\cal V}$.

\subsubsection{The geometric phase}
The operator ${\cal S}$ has an important geometric significance.
It  incorporates information about the geometric phase
\cite{Ber84,Sim83} that is associated to a history. To see this,
one has first to recall that a Hilbert space $H$ is a line bundle
over the projective Hilbert space $PH$, i.e. the equivalence class
of all vectors that differ by a multiplication with a complex
number. We shall denote an element of $PH$ as $[\psi]$.
 The inner product on $H$ inherits
two important geometric structures on $PH$: a metric
\begin{equation}
ds^2 = ||d| \psi \rangle||^2 - | \langle \psi|d| \psi \rangle |^2 ,
\end{equation}
and a $U(1)$ connection
\begin{equation}
A = -i \langle \psi |d| \psi \rangle .
\end{equation}
When a point of $PH$ evolves along a loop $\gamma$, its total phase change consists of a piece
 that depends upon the dynamics and a piece that is essentially the holonomy of the connection
$A$ \cite{Pag87, AnaAh87}. This is known as the Berry phase and equals
\begin{equation}
e^{i \theta_g[\gamma]} = e^{i \int_{\gamma} A} =
\exp \left( \int \langle \psi|d |\psi \rangle \right)
\end{equation}

The geometric phase can also be defined  for open paths. The trick is that  any path on the
 projective Hilbert space can be closed by joining its endpoints with a geodesic, with respect
 to the natural metric. The geometric phase of the loop thus constructed is then {\em defined} to
 equal  the geometric phase associated to the open path. Hence if $\gamma = [\psi( \cdot )]$ is
 a path on $PH$ its associated geometric phase can be found  \cite{SaBha88}
\begin{equation}
 e^{i \theta_g[\gamma]} = \exp \left(  \int_{t_i}^{t_f} dt \langle
\psi(t) | \dot{\psi}(t) \rangle \right) \langle \psi_i|\psi_f \rangle .
\end{equation}
This expression is defined only if the endpoints are not orthogonal.

Now let us consider a discretized approximation to an element $|
\psi ( \cdot )\rangle $ of ${\cal V}$. Let us write therefore,
\begin{equation}
\alpha =
 \otimes_{t_j} | \psi_{t_j} \rangle \langle \psi_{s_j} | ,
\end{equation}
 where $|\psi_{t_j}
 \rangle $
 are   normalized vectors
on $H_{t_j}$.

We then calculate
\begin{equation}
Tr \left( {\cal S} \alpha \right) = \langle \psi_{t_0} | \psi_{t_n}
\rangle
 \langle \psi_{t_1}|\psi_{t_0} \rangle \langle \psi_{t_2}|\psi_{t_1} \rangle
 \ldots \langle \psi_{t_{n}}| \psi_{t_{n-1}} \rangle  .
\end{equation}
Let us then assume that $ \max |t_j - t_{j-1}| = \delta t$, and we choose the number
of time steps $n$ very large, so that $\delta t \sim O(n^{-1})$.   Then $|\psi_{t_j} \rangle$
approximates a path $[\psi(t)]$ on $PH$. Writing formally $\alpha_{\psi( \cdot )}$ for the projector
 we get
\begin{eqnarray}
\log Tr \left( {\cal S} \alpha_{\psi( \cdot )} \right) = \log \langle \psi_{t_0}|
\psi_{t_n}
\rangle + \sum_{i=1}^n \log \langle \psi_{t_i} | \psi_{t_{i-1}} \rangle \nonumber
\\
= \log \langle \psi_{t_0}|\psi_{t_n} \rangle + \sum_{i=1}^n \log \left( 1 - \langle
\psi_{t_i}  | \psi_{t_i} - \psi_{t_{i-1}} \rangle \right) ,
\end{eqnarray}
which in  the limit of large $n$ yields
\begin{equation}
\log Tr \left( {\cal S} \alpha_{\psi( \cdot )} \right) = \log \langle
\psi_{t_0}|\psi_{t_n}
\rangle - \sum_{i=1}^n \langle \psi_{t_i} | \psi_{t_i} - \psi_{t_{i-1}} \rangle + O
((\delta t)^2).
\end{equation}
As $\delta t \rightarrow 0$ the sum in the right-hand side converges to a
Stieljes integral $ - \int_{t_i}^{t_f} dt \langle \psi(t)|\dot{\psi}(t)
\rangle $ and hence
\begin{equation}
Tr \left( {\cal S} \alpha_{\psi( \cdot )} \right) = e^{i \theta_{g}[\psi( \cdot )]}
\end{equation}
 This is  the Berry phase associated to the path $\psi( \cdot )$.
This implies that ${\cal S}$ exists as an operator on ${\cal V}$. Its matrix elements
can be defined as
\begin{equation}
\langle \phi( \cdot )|{\cal S} | \psi( \cdot ) \rangle = \langle \phi(t_0)| \psi(t_f) \rangle
\exp \left( \int \langle \psi(t)| d |\psi(t) \rangle \right),
\end{equation}
where the integral in the exponential is of the Stieljes type (rather than of the Lebesque,
that was
 used in the definition of $\otimes_{t \in T} H_t$). The Stieljes integral is  defined
for all measurable functions of bounded variation. Hence the
matrix elements of ${\cal S}$ are finite. This implies, it is a
well defined bounded operator and it is easy to check that it
remains unitary  even in the continuous limit.

\subsubsection{Another attempt to construct the decoherence functional}

We have showed that the main operators that form the decoherence functional exist in the continuous limit.
Could we then proceed and define a continuous-time decoherence functional from equation (2.20)?
The answer is no, at least not straightforwardly. The problem is that the analogue of the maps ${\cal A}$
does not exist in the continuous limit. The reason is the same as before: an embedding of single-time Hilbert spaces to the history Hilbert space fails to be continuous.

One has therefore two options. First, it should be noted that an
initial and final moment of time is necessary in the decoherence
functional, because they incorporate information about the
preparation of the system. From an operational viewpoint, one
could then say that the specification of an initial state cannot
be sharp in time and as such they ought to be incorporated in the
decoherence functional by an object that is extended in time. This
would imply a generalization of expression (2.23), where the map
${\cal A}$ is defined from ${\cal V}$ not to $\partial {\cal V}$,
but  to some other Hilbert space, which is associated with a
finite time sub-interval of $T$. The introduction of such an
operator could provide a construction of a
 continuous decoherence  functional in this case. This would be mathematically well-defined and operationally
meaningful, but would diverge from  the standard canonical quantum theory.
For this reason, we shall not pursue this further in this paper.

An alternative would be  to abandon the effort to define a continuous decoherence functional
and assume at most weak
continuity.

If we assume two one -dimensional projectors $\alpha_{\psi( \cdot )}$ and
$\alpha_{\psi( \cdot )}$ we get an expression for the decoherence functional with zero Hamiltonian
\begin{eqnarray}
d(\alpha_{\psi( \cdot )}, \alpha_{\psi( \cdot )} ) = \langle \psi(t_i) | \hat{\rho}_0| \psi(t_i) \rangle
\langle \psi(t_f) |   \psi(t_f) \rangle
\nonumber \\
\exp \left(   \int_{t_i}^{t_f} dt \langle \psi(t)|\dot{\psi}(t)
\rangle
 - \int_{t_i}^{t_f} dt \langle \dot{\psi(t)}|\psi(t)
\rangle \right).
\end{eqnarray}
In the special case where $\hat{\rho}_0 =  |\psi(t_i) \rangle \langle \psi(t_i) | =
| \psi(t_i) \rangle \langle \psi(t_i) | $ and
$  |\psi(t_f) \rangle \langle \psi(t_f) | = | \psi(t_f) \rangle
\langle \psi(t_f) | $  its
value is equal to
\begin{equation}
d(\alpha_{\psi( \cdot )}, \alpha_{\psi( \cdot )} ) = e^{i \theta_{g}[\psi( \cdot ), \psi( \cdot )] },
\end{equation}
the Berry phase for the {\it loop} formed from $\psi( \cdot )$ and $\psi( \cdot )$, since
now they have the same endpoints.

More interestingly, when the Hamiltonian is included the decoherence
functional becomes
\begin{equation}
d(\alpha_{\psi( \cdot )}, \alpha_{\psi( \cdot )} ) = \langle \psi(t_i) | \hat{\rho}_0| \psi(t_i) \rangle
\langle \psi(t_f) | \hat{\rho}_f | \psi(t_f) \rangle
e^{iS[\psi( \cdot )] - i S^*[\psi( \cdot )]},
\end{equation}
where the action is  given by the familiar expression (its variation gives
the Schr\"odinger equation)
\begin{eqnarray}
S[\psi( \cdot )] = \int_{t_i}^{t_f} dt \langle  \psi(t)|i \frac{d}{dt} - H|
\psi(t)\rangle .
\end{eqnarray}

One might then give equation (2.41) as a {\it definition} of a decoherence functional for
pairs of one-dimensional projectors and then extend this definition by finite addition to projectors
with finite trace. But there is no {\it a priori} guarantee that one would thus construct an object
taking finite values
a general projector on $\otimes_{t \in T} H_t$. Nonetheless, equation (2.41) {\em highlights the importance
of the action as the object relating kinematics, dynamics and the probabilistic structure of
quantum theory}.

We shall return to the issue of the definition of a continuous-time decoherence functional in
section 3.5.3.
\subsection{Time reversal}

A symmetry on a history Hilbert space is represented either by a unitary or an antiunitary
operator. This has been established by Schreckenberg \cite{Schr97} .

Of particular interest are the time reversal transformations. In discrete time they are defined
by \cite{IL95}
\begin{equation}
{\cal T} |v_{t_1} \rangle |v_{t_2} \rangle \ldots |v_{t_n} \rangle = |v_{t_n} \rangle |
v_{t_{n-1}} \rangle \ldots |v_{t_1} \rangle .
\end{equation}
Clearly
\begin{eqnarray}
{\cal T} {\cal T}^{\dagger} =1 \\
{\cal TST}^{\dagger} = {\cal S}^{\dagger}
\end{eqnarray}
Also for the operator ${\cal U}$ defined by (2.24) we have
\begin{equation}
{\cal TUT}^{\dagger} = e^{- i\hat{H} t_n} \otimes \ldots \otimes e^{-i\hat{H}t_1}
\end{equation}
and when time runs in the full real line
\begin{equation}
{\cal TUT}^{\dagger} = {\cal U}^{\dagger}
\end{equation}
Finally for the time inverted projection operators $ \alpha^T = {\cal T}\alpha {\cal
T}^{\dagger}$ (corresponding to homogeneous histories) we have
\begin{equation}
d(\alpha^T, \beta^T) = Tr ( \hat{C}^{T \dagger}_{\alpha} \rho_0 \hat{C}^T_{\beta}
\rho_f)
\end{equation}
where $\hat{C}^T_{\alpha} = \hat{\alpha}_{t_1}(t_n) \ldots \hat{\alpha}_{t_n}(t_1)$. In the discrete case
this form is not transparent,
but when time takes values in all ${\bf R}$ the Heisenberg picture operators
transform as  $\alpha_t(t) \rightarrow \alpha_t(-t) $ and therefore
\begin{eqnarray}
d(\alpha^T, \beta^T) = d(\beta, \alpha) = [d(\alpha,
\beta)]^*
\end{eqnarray}
Of course this later equation does not hold if  the Hamiltonian is
time -dependent and the system is not time-homogeneous.

The operator ${\cal T}$ is naturally defined also on $\otimes_{t \in{\bf R}}H_t$
\begin{equation}
{\cal T} | \psi( \cdot ) \rangle = {\cal T} | \psi (- \cdot)
\rangle
\end{equation}

It is important to note that the time reversal operator is linear rather
than anti - linear as in canonical quantum mechanics. This has again to do
with the presence of two laws of time transformations in history theories;
here time reversal implemented by ${\cal T}$ corresponds to the causal,
kinematical properties of time. The time
inversion operator of canonical quantum mechanics is obtained by the study of
the Schr\"odinger equation and as such is clearly associated to the dynamical
aspect of time.

Of course we can always define  an anti-linear time reversal
operator in complete analogy with the canonical case; a complex
conjugation on $H$ naturally defines a complex conjugation on
${\cal V}$. It would act on the Heisenberg picture operators as
$\alpha_t(t) \rightarrow \alpha_t(-t)$.

\subsection{Summary}
Let us summarize here the results of this section. We showed how a
Hilbert space $\otimes_{t \in T}H_t$ for continuous time histories
can be constructed and how time-averaged observables can be
defined as operators acting on it. Then we argued that in general
 we will have to restrict to a particular subset of $\otimes_{t \in T}H_t$.
 We then showed that the decoherence functional
cannot be defined as a limiting case of its discrete-time form.

We then analyzed the structure of the decoherence functional. We
identified the pieces out of which it is constructed, in light of
the two laws of time transformation of history theories, and
showed their relation to the dynamical and geometric phase of
canonical quantum theory. We discussed a possible way to construct
the continuous-time decoherence functional and finally saw how
{\it unitary} time-reversal transformations are implemented in
this scheme.

\section{Phase space histories}

In the previous section we  examined the general structure of continuous-time histories, without
making any reference to a particular physical system, or class of systems.
In order to do so, we  necessarily have to make reference to a corresponding classical system and seeks to
 identify operators on the Hilbert space with observables that have a classical analogue. This
is, in effect, the {\it quantization} procedure. In this section,
we will study how the classical phase space structure is
manifested in the histories formalism.

We refer the reader
to section 1.3.1 for explanation of the notations we will use in this section.

\subsection{The canonical group}
\subsubsection{The Weyl group}
In quantum theory the information about the corresponding classical theory can be
 encoded in the  {\it canonical group}. This is
 classically identified as a group that acts transitively by canonical transformations
  on the classical phase space
$\Gamma$ \cite{I83}. When $\Gamma = {\bf R}^{2n}$
 the canonical group is the $(2n+1)$- dimensional Weyl group. This   is defined
 whenever the phase space has a vector space structure. It can therefore be
infinite dimensional, as in a field theory.
 For its definition  an inner product on $\Gamma$ has to be assumed, so we usually consider
$\Gamma$ to be a real Hilbert space.

The Weyl group  is generated by $q_i,p_i,1$ and has basic Lie
algebra relations
\begin{eqnarray}
\{q_i,q_j \} = 0, \\
\{ p_i, p_j \} = 0, \\
\{q_i, p_j \} = \delta_{ij}.
\end{eqnarray}
A  generator of the Weyl group reads  $\chi \cdot p + \xi \cdot q
+ c$, in terms of the inner product in $\Gamma$, and is labelled
by $(\chi_i, \xi_i,c)$.
 The  corresponding group  element will be denoted as $(\chi,\xi,c)$.
The group multiplication law is
\begin{equation}
(\chi_1,\xi_1,c_1) \cdot (\chi_2,\xi_2,c_2) = (\chi_1+\chi_2, \xi_1 +\xi_2, c_1 + c_2 +
\frac{1}{2}(\xi_1 \cdot \chi_2 - \xi_2 \cdot \chi_1)) .
\end{equation}
When the canonical group has been identified, the Hilbert space of the theory is constructed through the selection of
 one of its  unitary  {\it irreducible} representations. The criterion for this selection is  the existence of
self-adjoint operators that  correspond to the generators of  classical symmetries
(e.g. the Hamiltonian, the Lorentz group etc).

\subsubsection{Coherent states}
Suppose we have a representation of the canonical group by unitary operators
$\hat{U}(g)$ on a Hilbert space. Furthermore, let $\hat{h}$ denote the Hamiltonian of this system and by  $|0 \rangle_H$ the vacuum,
i.e. the  Hamiltonian's lowest eigenstate. Then we define the coherent states as the vectors
\begin{equation}
|g \rangle = \hat{U}(g) | 0 \rangle .
\end{equation}
Now consider the equivalence relation on the canonical group defined as $g \sim g' $ if $|g \rangle$
and $|g' \rangle$ correspond to the same ray. The phase space $\Gamma$ is identified as the quotient space $G/ \sim $
 and we can label a coherent state by points $z \in \Gamma$.

Hence the canonical group defines a map $i: \Gamma \rightarrow PH $ as $ z \rightarrow | z \rangle$. As we explained $PH$ has a natural metric and a U(1) bundle structure with a connection. These structures  can be pullbacked to $\Gamma$ with $i^*$. We have then on $\Gamma$ a $U(1)$ bundle with a connection $A$ given by  \begin{equation}
A = -i \langle z|d| z \rangle
\end{equation} and a metric
\begin{equation}
ds^2 = || d | z \rangle ||^2 - | \langle z |d| z \rangle|^2 ,
\end{equation}
where $d$ is the exterior derivative on $\Gamma$.
 The fundamental property of
  coherent states is that they are an overcomplete basis; i.e.
  any vector $|\Psi \rangle$ can be written as
\begin{equation}
|\Psi \rangle = \int d \mu(z) f(z) |z \rangle ,
\end{equation}
in terms of some complex-valued function $f$ on $\Gamma$.
Here $d \mu$ denotes some
 natural measure on $\Gamma$. In $n$ dimensions it is equals $\frac{d^n\bar{z} dz}{(2 \pi)^n}$.
There is also a decomposition of the unity
\begin{equation}
\int d \mu(z) |z \rangle \langle z| = \hat{1} .
\end{equation}
If the phase space $\Gamma$ has a vector space structure the canonical group is the Weyl group.
Its most usual representation is on $e^{\Gamma_C} =  \oplus_{n=1}^{\infty} (\otimes_n H)_S$, the symmetric Fock space generated by
 the complex vector space  $\Gamma_C$,  a complexification of $\Gamma$ \cite{Ber66, Wald94}.
 On the Fock space there exist the unnormalized coherent states
$| \exp  z \rangle $ that  to each $z \in \Gamma_C$ they assign the vector $| \exp z \rangle = \oplus_{n=0}^{\infty} \otimes_n z $.
The inner product of such states is given by
\begin{equation} \langle \exp z'| \exp z \rangle = e^{(z',z)_C} ,
\end{equation} where  $(,)_C$ denotes an inner product on $\Gamma_C$
(its choice depends upon the way $\Gamma$ is complexified). The
corresponding normalized states  are denoted simply as $| z
\rangle$, or $|\chi,\xi \rangle$.

 \subsubsection{The overlap kernel}
For the finite dimensional Weyl group, the Stone-von Neumann theorem asserts that all
irreducible representations are unitarily equivalent  to the Fock one. This is not true in
 infinite dimensions. In this case, the information about the representation is encoded in
 the coherent states  overlap  $ \langle \chi' \xi'| \chi \xi \rangle$.

This is determined by the expectation functional $K(\chi,\xi) = \langle 0 | \chi, \xi \rangle$
as a consequence of the group combination law \begin{equation}
 \langle \chi' \xi'|\chi  \xi \rangle  = e^{i/2 (\chi \cdot\xi' - \xi \cdot \chi')} K(\chi-\chi',\xi - \xi').
\end{equation}
The expectation functional suffices to describe the connection and metric structure on phase  space. If we write $K = e^W$, we find
\begin{eqnarray}
A = \xi_i d\chi^i , \\
ds^2 = - Re \left(  \frac{\partial^2W}{\partial \chi^i \partial \chi^j} d\chi^i d\chi^j
+ \frac{\partial^2 W}{\partial \xi^i \partial \xi^j} d \xi^i d \xi^j \right. \nonumber \\
\left.
+  (\frac{\partial^2 W}{ \partial \chi^i \partial \xi^j} + \frac{\partial^2W}{\partial \chi^j \partial
\xi^i}) d\chi^i d \xi^j \right).
\end{eqnarray}

In the case of an harmonic oscillator with frequency $\omega$, the
functional $W$ reads \begin{equation} W(\chi,\xi) =
-\frac{1}{2}[\omega \chi^2 + \omega^{-1} \xi^2] . \end{equation}

The knowledge of the overlap suffices to construct the Hilbert space and the representation
\cite{Kla85}.

A vector of the Hilbert space can be constructed as a function on phase space of the form
$\Psi(\chi, \xi) = \sum_l c_l \langle \chi  \xi| \chi_l  \xi_l \rangle $ for a
finite number of complex  numbers $c_l$ and $\chi_l, \xi_l$. The inner product between two vectors characterised by
$c_l,\chi_l, \xi_l$ and $c'_l,\chi'_l, \xi'_l$ is
\begin{eqnarray}
 \sum_l c'^*_l c_l \langle \chi'_l \xi'_l|\chi_l  \xi_l \rangle .
\end{eqnarray}
The Weyl group is then represented by the operators $\hat{U}(\chi,\xi)$, which are defined as
\begin{equation}
(\hat{U}(\chi',\xi') \Psi)(\chi,\xi) = e^{\frac{i}{2}(\chi' \cdot \xi - \xi' \cdot \chi)} \Psi(\chi-\chi',\xi-\xi') .
\end{equation}

The above is written for the finite-dimensional Weyl group, but with little modification
is also valid for the infinite dimensional case. The only difference is that in finite
dimensions the Stone- von Neumann theorem holds: {\it all irreducible, strongly continuous,
 unitary representations of the Weyl group, are unitarily equivalent}.

In the infinite dimensional case the vector space out of which the
Weyl group is constructed is a functional space. For field
theories in Minkowski spacetime this is a subspace of the space of
square integrable functions on ${\bf R}^3$. In this case, the
group of spatial translations is also represented unitarily on the
Hilbert space. If the vacuum is the {\it unique translationary
invariant state} in the representing Hilbert space, then it can be
proven that  {\it all unitarily equivalent representations share
the same expectation functional}, and conversely,
 if two representations differ in their expectation functionals, they
  are unitarily inequivalent \cite{Ara60}.
We shall refer to this as the {\it uniqueness theorem} for the expectation functional.

\subsection{Classical histories}
In order to study the phase space structure of quantum mechanical
histories, we need to describe histories in classical mechanics in
a way that is amenable to a direct comparison. We shall,
therefore, reproduce here the main points of this description,
referring the reader to \cite{Sav99b, SavAn00} for details.

Consider the space of classical histories $\Pi$ viewed as the set of
continuous paths on the classical phase space $\Gamma$. An element of
$\Pi$ is a path $\gamma:T \rightarrow \Gamma$.

For any function $f$ on $\Gamma$ one can define a family of functions $F_t $ on $\Pi$ as
\begin{equation}
F_t(\gamma) = f(\gamma(t)). \end{equation} Taking for simplicity $\Gamma = {\bf R} \times {\bf R} = \{ (q,p) \}$, we can define  $q_t$ and $p_t$ as elements of $C^{\infty}(\Pi)$ through \begin{eqnarray} q_t (\gamma) = q(\gamma(t)), \\ p_t(\gamma) = p(\gamma(t)). \end{eqnarray} Two other functions on $\Pi$ can be identified \begin{eqnarray} V(\gamma) = \int_T dt p_t \dot{q}_t (\gamma), \\ H(\gamma) = \int_T dt h(p_t,q_t) , \end{eqnarray} with $h$ denoting the standard canonical Hamiltonian. If we furthermore equip $\Pi$ with a symplectic form  \begin{equation} \omega = \int dt dp_t \wedge dq_t , \end{equation} corresponding to the
Poisson bracket
\begin{equation}
\{ q_t,p_{t'} \} = \delta(t,t'),
\end{equation}
we can examine the canonical transformations generated by the
functions $V$ and $H$. These are the generators of the two
distinct laws of time transformation that characterize history
theories.

The transformations  generated by $V$ perform translations of the
$t$ argument in a path, that is $\gamma \rightarrow \gamma'$ with
$\gamma'(t) = \gamma(t+s)$ (s the affine parameter of the
corresponding one-parameter group). Or in its induced action on
the functions
\begin{equation}
F_t \rightarrow F_{t+s}.
\end{equation}
$H$ respects the time labelling of the points of the path.  It
acts on each point of the path by transforming it (while keeping t
fixed) according to Hamilton's equations. This means its action on
$C^{\infty}(\Pi)$ is
\begin{equation}
(q_t,p_t) \rightarrow (q_t(s),p_t(s)),
\end{equation}
where $q_t(s)$ is the function that to each path $\gamma$ assigns
the $q$- coordinate of the point obtained by integrating the
Hamilton equations from initial point with coordinates $(q_t,p_t)$
to time $s$; similarly for $p_t(s)$.

In the classical setting this distinction of two laws of time
transformation, is nicely related to the least action principle. A
path $\gamma$ is a solution to the classical equations of motions
iff it is a fixed point of the canonical transformation generated
by the action $S = V - H$. This implies the condition
\begin{equation}
\{q_t, S \}(\gamma) = \{p_t,S \}(\gamma) = 0.
\end{equation}
Hence for the solutions to the equations of motion  the laws of time evolution
generated by $V$ and
$H$ coincide.

\subsection{The history group}
The construction of the history Hilbert space through the tensor product of single-time Hilbert spaces suggests a natural generalisation; the history Hilbert space has to carry the representation of
the history group, the history analogue of the canonical group \cite{IL95}. This is a group that acts by symplectic transformations on the
space of phase space histories.
For linear phase spaces  this is
\begin{eqnarray}
 [q^ i_t,p^j_{t'}] = i \delta^{ij} \delta(t,t'),
 \end{eqnarray}
It is clearly an infinite dimensional Weyl group. Its  proper
definition involves a choice of smearing functions: we define
$q_{\xi} = \int d \mu(t) \xi_i(t) q^i_t$ and $p_{\chi}$ similarly,
and write the commutator as

\begin{equation}
 [q_{\xi},p_{\chi}] = i  \int d \mu(t) \chi(t) \cdot \xi(t)
\end{equation}
The precise choice of a test-function space depends on the physics of the system,
but it definitely has  to consist of square-integrable functions,
 if the right-hand-side of (3.28)
is to be defined. Here $d \mu$ stands for any measure
on the real line, but what is mainly used is the measure employed in the construction of
$\otimes_t H_t$, i.e. $d \mu(t) = dt / \tau$.

 This history group is an infinite dimensional Weyl group and admits many
unitarily inequivalent representations.

The analysis of the classical histories suggests the criterion for selecting a representations. {\it There should exist self-adjoint operators in the Hilbert space, that correspond to the functions $V$ and $H$ of the classical theory}.
 For quadratic Hamiltonians, a Fock representation
(that has the structure of a continuous tensor product)
can be constructed \cite{ILSS98}, in which both the Hamiltonian $H_{\kappa}$ and an operator corresponding to $V$ (the Liouville operator) exist. An important feature of this construction is the existence of a Hilbert space vector $| 0 \rangle$, which is the lowest eigenstate of the Hamiltonian and is left invariant under the action of $e^{isV}$ \cite{ILSS98,Sav99a}.
The  projector $|0 \rangle \langle 0 |$ corresponds to the proposition that at all times the systems is to be found in the ground state.

Another important feature of this construction is the fact that
{\em the continuous tensor product of coherent states of the
harmonic oscillator exists as a coherent state in the Fock Hilbert
space}. This is a feature that can be generalized for systems with
non-quadratic Hamiltonian. Indeed, it will form the basis of our
construction.

\subsubsection{General representations}
Representations cannot be explicitly constructed for non - quadratic
Hamiltonians
(it is the same situation with the one  in canonical quantum field theory). Nonetheless, if we have some information
about the canonical theory,  we can exploit this to construct representations for the
history group.

As we explained in section 2.3, unbounded operators can be defined
on a history Hilbert space, if we start our construction from a
subset ${\cal L}$ of the Hilbert space. Since we want a Hilbert
space that carries a representation of the history group, the
natural choice for ${\cal L}$ would be the coherent states of the
corresponding canonical group. If  $H$ carries a representation of
the canonical group  $\hat{U}(\chi,\xi)$ and $\hat{h}$ is the
Hamiltonian with a unique ground state $|0 \rangle_H$, we define
the canonical coherent states $|z \rangle = |\chi\xi \rangle =
\hat{U}(\chi, \xi) |0 \rangle_H$. Then  the history Hilbert space
${\cal V}$ is  generated by all vectors
\begin{equation}
| z( \cdot ) \rangle = |\chi( \cdot ) \xi( \cdot ) \rangle := \otimes_{t \in T} |\chi_t \xi_t \rangle_{H_t}
\end{equation}
 Furthermore, we  demand that the vectors $|z( \cdot ) \rangle $ on ${\cal V}$ are
 the coherent  states associated with the corresponding history group.
 In this case we shall have a history  overlap kernel
\begin{equation}
\langle \chi'( \cdot ) \xi'( \cdot ) | \chi( \cdot ) \xi( \cdot ) \rangle = \exp \left( \int d \mu(t)  \log ( \langle \chi'_t \xi'_t|\chi_t \xi_t \rangle_{H_t} ) \right) .
\end{equation}
The corresponding expectation functional $K_h[\chi( \cdot ),\xi( \cdot )] = e^{W_h[\chi( \cdot ),\xi( \cdot )]}$  will
 read in terms of the canonical expectation
functional $K(\chi, \xi)= e^{W[\chi,\xi]}$
\begin{equation}
W_h[\chi( \cdot ),\xi( \cdot )] = \int d \mu(t) W[\chi_t,\xi_t] .
\end{equation} Clearly certain conditions have to be imposed on the admissible paths $(\chi,\xi)( \cdot )$ if the  integral is to be finite. (We shall take $T = {\bf R}$ without any loss of generality in this
section.)

Now,  there exists a norm $|\cdot |_{\Gamma}$ on the phase space
( it can be constructed  from the metric (3.7) or from the inner
product). This induces a norm in the
space of paths $t \rightarrow z_t$, which is  given by
\begin{equation}
|z( \cdot ) |_{\Pi} = \int d \mu(t) |z_t|_{\Gamma} .
\end{equation}
Our first restriction, will be to consider  only continuous paths with a  finite value for the norm.
  For simplicity we shall  assume that the maps
$z( \cdot )$ take  values $(0,0)$ except within compact subsets of
${\bf R}$. But we expect that our results would still be valid, if
the paths $x(\cdot)$  converge to $(0,0)$ sufficiently fast
(exponentially) outside compact sets.

We shall also assume that the canonical coherent states, viewed as
maps from the phase space to $H$ are  smooth functions of their
arguments. This implies that   $W[\chi,\xi]$ is a smooth function
of its variables. Since by definition   $W[0,0] = 0$, the above
conditions are sufficient for the integral (3.31) to be finite.

We shall also impose the restriction
that the maps $z( \cdot )$ are everywhere {\it Lifschitz}: in any compact subset $U$ of ${\bf R}$, there
exists $C > 0$, such that for all $t_1,t_2 \in U$, $|z_{t_1} - z_{t_2}|_{\Gamma} < C |t_1 - t_2|$. This is a stronger assumption than continuity, but weaker than differentiability and it is necessary for proving existence of the Liouville operator.

If $| \chi( \cdot ) \xi( \cdot ) \rangle$ are to correspond to coherent states, {\it they have to be continuous
functions of their arguments}. This is proven as follows:
\\ \\
Let us assume that $|z_1( \cdot ) - z_2( \cdot ) |_{\Pi} = \delta >0$. Then
\begin{eqnarray} || |z_1 ( \cdot )\rangle - |z_2( \cdot ) \rangle ||^2 = 2 ( 1 - \cosh \int d \mu(t) \log \langle z_{1t}|z_{2t} \rangle) .
\end{eqnarray}
 Let us write $| f_t \rangle = |z_{2t} \rangle - |z_{1t} \rangle $.    Then we have
\begin{eqnarray}
|| |z_1 ( \cdot )\rangle - |z_2( \cdot ) \rangle ||^2_{{\cal V}} = 2 - 2 \cosh \int d \mu(t) \log (1 + \langle z_t|f_t \rangle) . \end{eqnarray} The finiteness of $||z_1( \cdot ) - z_2( \cdot )||_{\Pi}$, implies that except for a set  of measure zero, there
exists  $c \geq 0 $, such that $|\langle z_t|f_t \rangle| \leq \sqrt \langle f_t|f_t \rangle < c \delta$. Now, there exist complex numbers $c_t$,
such that $\log (1 + \langle z_t  |f_t \rangle) = c_t\langle z_t|f_t \rangle$. By our previous result (except perhaps in a set of measure zero)
these $c_t$ satisfy $|c_t| < C$, for  $C >0$. Using this result, we get
\begin{eqnarray} || |z_1 ( \cdot )\rangle - |z_2( \cdot ) \rangle ||^2_{{\cal V}} = 2  - 2 \cosh \left( \int d \mu(t) c_t \langle z_t|f_t \rangle \right).
\end{eqnarray}
The integral is bounded $| \int d \mu(t) c_t \langle z_t|f_t \rangle | \leq C \delta$, so for sufficiently small $\delta$, there
exists a constant $C' >0$ such that
 \begin{eqnarray}
|| |z_1 ( \cdot )\rangle - |z_2( \cdot ) \rangle ||^2_{{\cal V}} \leq C' \delta^2 ,
\end{eqnarray}
showing that $|z( \cdot ) \rangle$ is a continuous function of $z( \cdot )$.
\\ \\
This implies that $W$ is also a continuous function of $\chi( \cdot ), \xi( \cdot )$; so as explained in section
3.1.3, we  define a representation of the history group using equation (3.16).

But the representation can also be defined straightforwardly. Indeed,
we can write a unitary operator $U(\chi( \cdot ),\xi( \cdot ))$ as $\otimes_{t \in {\bf R}} U(\chi_t,\xi_t)$, i.e.
by its action on the coherent state   vectors
\begin{equation}
 U(\chi( \cdot ),\xi( \cdot )) | \chi'( \cdot ) \xi'( \cdot ) \rangle =
e^{\frac{i}{2} \int d \mu(t) (\chi'_t \cdot \xi_t-\chi_t \cdot
\xi'_t)} |\chi( \cdot ) +\chi'( \cdot ), \xi( \cdot ) + \xi'(
\cdot ) \rangle .
\end{equation}
 Therefore time averaged operators for position $q_{\xi} = \int d \mu(t) q_t \cdot \xi(t)$ and momentum $p_\chi = \int d \mu(t) p_t \chi(t) $
{\em do exist} on ${\cal V}$.

\subsubsection{Operators}
Let us first see how we can define the analogue of the Hamiltonian  $H_{\kappa} = \int d \mu(t)
h(q_t,p_t)$
 in this Hilbert space.
\\ \\
Let $\hat{h}$ be the Hamiltonian on the canonical Hilbert space. We assume that the representation
of the Weyl group can be chosen, so that all
 coherent state vectors lie in the domain of $\hat{h}$. This suffices to show that there exist
complex numbers
$A(s)$, such that
\begin{equation}
 \langle \chi' \xi'|e^{-i\hat{h}s}  |\chi \xi \rangle  =
  \langle \chi' \xi'| \chi \xi \rangle ( 1 -i A(s) h(\chi,\xi;\chi',\xi')  s),
\end{equation}
 where $ h(\chi,\xi;\chi',\xi') =
 \langle \chi' \xi'|\hat{h}|\chi \xi \rangle /\langle \chi' \xi'|\chi \xi \rangle$,
and for each neighborhood of $s=0$ there exists $C >0$ such that $|A(s)| \leq C$.
 Let us try to define a version of the operator $U_{\kappa}(s) = e^{-iH_{\kappa}s}$
as $\otimes_t e^{-i \hat{h} \kappa(t) s}$. It is easy to show, as in section 2.3.3,
 that it is   well-defined; the issue is to show it is continuous
 at $s = 0$, for then by Stone's theorem $H_{\kappa}$ exists. We have
\begin{eqnarray} |\langle \chi'( \cdot ) \xi'( \cdot )|U(s) - 1|\chi( \cdot )
 \xi( \cdot ) \rangle |   \nonumber \\ = |\exp \left( \int d \mu(t) \log (
 \langle \chi'_t \xi'_t|e^{-i\hat{h}\kappa(t)s}|\chi_t \xi_t \rangle_{H_t} ) \right)
- 1 |  \nonumber \\
= |\exp \left( \int d \mu(t) \log ( 1 - i A(s)s \kappa(t)
 h(\chi_t,\xi_t;\chi'_t, \xi'_t)) \right) - 1|
\nonumber \\ \leq C |\int d \mu(t) \kappa(t) h(\chi_t,\xi_t:\chi'_t, \xi'_t)| |s|
 .
 \end{eqnarray}

 Here $C$ is a real positive number. $U(s)$ has therefore  matrix  elements
 continuous with respect to $s$ if $\int d \mu(t) \kappa(t) h(\chi_t, \xi_t;
\chi'_t,\xi'_t)$ exists. We can take $\kappa(t)$ to be a  measurable function
that grows at most  polynomially. If we have adjusted $\hat{h}$ so  that
$\hat{h} | 0 \rangle_H = 0$, then it suffices that $h(\chi_t,\xi_t;\chi'_t,\xi'_t)$
 is continuous. For we have demanded that $(\chi_t, \xi_t) \rightarrow 0$
exponentially fast outside some compact set, hence
$h(\chi_t,\xi_t;\chi'_t, \xi_t') \rightarrow 0$ outside this
compact set.

  The operator $H_{\kappa}$
 {\em can be therefore defined}.  \\ \\
A Liouville operator corresponding to  the classical function
 $\int dt p_t \dot{q}_t$ is also  defined by its action on coherent states
\begin{equation}
e^{isV} | \chi( \cdot ) \xi( \cdot ) \rangle \rightarrow =
| \chi'( \cdot ) \xi'( \cdot ) \rangle ,
\end{equation}
where $(\chi'(t),\xi'(t)) = (\chi(t+s), \xi(t+s))$. We need to
check that it is continuous at $s=0$. We have
\begin{equation}
\langle \chi( \cdot ) \xi( \cdot )|e^{isV}|\chi( \cdot ) \xi( \cdot )
 \rangle = \exp \left( \int d \mu(t) \log \langle \chi_t \xi_t| \chi_{t+s}
 \xi_{t+s} \rangle_{H_t} \right).
 \end{equation}
Since the coherent states are continuous functions of their
arguments and the paths  $(\chi, \xi)( \cdot )$ have been assumed
Lifschitz, there exists a vector $|f_t,s \rangle_{H_t}$ on $H$
such that
\begin{equation}
 |\chi_{t+s} \xi_{t+s} \rangle_{H_t} = |\chi_t \xi_t \rangle_{H_t} + s |f_t,s \rangle_{H_t} ,
\end{equation}
 and  $\langle f_t,s|f_t,s \rangle < C_t $ for some constants $ C_t > 0$.
Therefore
\begin{eqnarray}
|\langle \chi( \cdot ) \xi( \cdot )|e^{isV} - 1|\chi( \cdot ) \xi( \cdot ) \rangle |
\nonumber \\
= |\exp \left( \int d \mu(t)
\log ( 1 + s \langle \chi_t \xi_t|f_t,s \rangle)_{H_t} \right) - 1 |\nonumber \\
%= | \exp \left( A s \int d \mu(t)   \langle \chi_t \xi_t|f_t,s \rangle_{H_t} \right) - 1|
%\nonumber \\ \leq A |s| \int d \mu(t) |\langle \chi_t \xi_t|f_t,s \rangle |
\leq A |s| \int d \mu(t) C_t,
\end{eqnarray}
for some constant $A > 0 $. Now, since we assume $(\chi_t,
\xi_t)\rightarrow (0,0)$   outside compact intervals, $C_t$ can
always be chosen to be constant in this compact interval  and
vanish outside this, thus rendering the integral finite. We
therefore establish continuity of the matrix elements of $e^{isV}$
around $s=0$.

The operator $V$, therefore, exists.
\\ \\
The existence of $V$ and $H_{\kappa}$
 also implies the existence of an action operator $S_{\kappa} = V - H_{\kappa}$.
\\ \\
To summarize,  assuming that: \\
1. the canonical coherent states are smooth functions of their arguments, \\
2. they lie in the domain of $\hat{h}$, \\
3. $\hat{h}$ has a unique ground state $|0 \rangle_H$, in which $\hat{h} |0 \rangle_H = 0$ \\
4. we consider paths $t \rightarrow z_t$, that  satisfy the Lifschitz condition, \\  \\
we can define a representation of the history group in a Hilbert space ${\cal V}$ in the fashion described, such that  the two generators of time-transformation are self-adjoint operators on ${\cal V}$ .

An issue that can be raised at this point is that the choice of
paths is restricted to ones that $(\chi_t,\xi_t) \rightarrow 0$ as
time goes to infinity. These are not sufficient to describe all
conceivable phase space motions, as for instance this  of
oscillators that oscillate eternally. However, one can consider
such properties in an arbitrarily large --but finite-- time
interval. This restriction is a consequence of the way we have
chosen to define the smearing functions for the generators of the
history group. A possibility that might lift this difficulty in a
natural manner is briefly presented in section 4.1.

\subsubsection{Uniqueness of the representation}
As $(\chi_t, \xi_t) \rightarrow (0,0)$ for large $t$,  the only vector that is left invariant
under the time translations generated by the Liouville operator is the ``vacuum'' vector $|0 \rangle
= \otimes_{t \in {\bf R}} |0 \rangle_{H_t}$. (It corresponds to the proposition that the system is on the ground state at all times).
Since the history Weyl group is isomorphic to the Weyl group
of a field theory, we can use the uniqueness theorem for the expectation functional, to establish
that any two of the representations, we have constructed are unitarily
inequivalent, if they have {\it different expectation functionals}.

This has different implications according to whether the {\it
canonical} Weyl group is finite or infinite dimensional. If it is
infinite dimensional and corresponds to a well behaved quantum
field theory (i.e. with a unique translationally invariant
vacuum), then the expectation functional of the canonical theory
is independent of the representation and unique. Hence, the
expectation functional for the history theory,  constructed by
equation (3.30) is also unique. This means for a given
representation of the canonical group, we can obtain a
representation of the history group, in such a way, that unitarily
equivalent representations of the canonical group yield unitarily
equivalent representations of the history group. This is, indeed,
very satisfactory.

But for finite dimensional canonical Weyl group, all
representations are unitarily equivalent. Hence different
expectation functionals correspond to unitarily equivalent
theories. But different expectation functionals canonically, lead
to different expectation functionals for the history group. And
these  give rise to unitarily inequivalent representations. We are
then in the unpleasant situation of having many inequivalent
history theories corresponding to one canonical theory. There is
no remedy for this. But, we should remark that the conditions
developed throughout this section, constrain severely the choice
of the representation of the canonical group, we are allowed to
use. The canonical coherent states have to lie in the domain of
all operators that  we want to also define in the histories
theory. Even if this does not guarantee uniqueness, at least it
gives a guideline for which type of representations are
interesting to use.

\subsubsection{The decoherence functional}

We saw that we have to restrict to paths $(\chi_t,\xi_t)$ that fall to zero rapidly at large $t$.
This means that the single-time Hilbert space at $t = \pm \infty$ is essentially one dimensional,
consisting  only of the vector $| 0 \rangle$.

We saw that in the construction of the decoherence functional, the
main problem came from the operators defined at the boundary
Hilbert space. In this construction, when time is taken in the
whole of the  real line, the boundary Hilbert space is
one-dimensional and the boundary operator ${\cal A}$ is just
multiplicative. Hence the decoherence functional splits in the
product of two phases:
\begin{equation}
d(\alpha, \beta) = Tr_{{\cal V}}( {\cal SU}^{\dagger} \alpha {\cal U}) Tr_{{\cal V}} ( {\cal S}^{\dagger}
{\cal U}^{\dagger} \beta {\cal U})   \end{equation}  The operator ${\cal U} $ is easily identified as $e^{-i H_{\kappa}}$ for $\kappa(t) = t$.

The  construction of the operator ${\cal S}$ is more intricate. Complex
analyticity of the coherent states makes consideration of the
diagonal matrix elements sufficient.

From the basic operation of the Weyl group we get that
\begin{equation}
\langle \chi' \xi' |\chi \xi \rangle = \exp \left( i/2 (\xi \cdot \chi' - \chi \cdot \xi') + W[\chi-\chi', \xi-\xi'] \right)
\end{equation}
 Assuming a discetization $ t_0 , t_1, \ldots t_n = t_f$  of the  interval $[t_i, t_f]$ the definition (2.22)
yields
\begin{eqnarray}
\langle \chi_{t_0},\xi_{t_0}; \ldots; \chi_{t_n}, \xi_{t_n}  | {\cal S} | \chi_{t_0} ,\xi_{t_0} ; \ldots \chi_{t_n}, \xi_{t_n} \rangle
= \langle \chi_{t_0} \xi_{t_0}|\chi_{t_n} \xi_{t_n} \rangle \prod_i \langle
\chi_{t_i} \xi_{t_i} | \chi_{t_{i-1}} \xi_{t_{i-1}} \rangle \nonumber \\ = e^{ i/2 (\xi_{t_n} \cdot \chi_{t_0} - \chi_{t_n} \cdot \xi_{t_0}) + W[\chi_{t_n} - \chi_{t_0},
\xi_{t_n} - \xi_{t_0}] } \nonumber \\
\times
\exp \left( \sum_i \frac{i}{2} ( \xi_{t_{i-1}} \cdot \chi_{t_i} -  \xi_{t_i} \cdot \chi_{t_{i-1}}) +
W[\chi_{t_{i-1}} - \chi_{t_i}, \xi_{t_{i-1}} - \xi_{t_i} ] \right) \nonumber \\ =  e^{ \frac{i}{2} ( \xi_{t_n} \cdot \chi_{t_0} - \chi_{t_n} \cdot \xi_{t_0}) + W[\chi_{t_n} - \chi_{t_0}, \xi_{t_n} - \xi_{t_0}] } \nonumber \\
\times
 \exp \left( \sum_i \frac{i}{2} [ \xi_{t_i} \cdot (\chi_{t_i}  - \chi_{t_{i-1}}) -
\chi_{t_i} \cdot (\xi_{t_i} - \xi_{t_{i-1}})] \right. \nonumber \\
\left.  - \frac{\partial W}{\partial \xi}[\chi_{t_i},\xi_{t_i} ]  (\xi_{t_i} - \xi_{t_{i-1}}) - \frac{\partial W}{\partial \chi}[ \chi_{t_i}, \xi_{t_i}] (\chi_{t_i} - \chi_{t_{i-1}} ) \right)    \end{eqnarray} Hence at the continuous limit get
\begin{eqnarray}
\langle \chi( \cdot ) \xi( \cdot )| {\cal S} | \chi( \cdot ) \xi( \cdot ) \rangle =
 e^{ \frac{i}{2} ( \xi(t_f) \cdot \chi(t_0) - x(t_f) \cdot \xi(t_0)) +
  W[\chi(t_f) - \chi(t_0), \xi(t_f) - \xi(t_0)] } \nonumber \\
\times
 \exp \left(  \int_{t_0}^{t_f} dt
\frac{i}{2} (\xi \cdot \dot{\chi} - \chi \cdot \dot{\xi} ) -
\int_{W_{t_0}}^{W_{t_f}}  dW
\right) = \nonumber \\
 \exp \left( \frac{i}{2} ( \xi(t_f) \cdot \chi(t_0) - \chi(t_f) \cdot \xi(t_0)) +
 W[\chi(t_f) -  \chi(t_0), \xi(t_f) - \xi(t_0)] \right. \nonumber \\
\left.
- W [\chi(t_f), \xi(t_f)] + W[ \chi(t_0) , \xi(t_0) ] \right) \nonumber \\
\times \exp \left(  \frac{i}{2} \int_{t_0}^{t_f} dt (\xi \cdot
\dot{\chi} - \chi \cdot \dot{\xi}) \right)
\end{eqnarray}
Clearly as $[t_0,t_f] \rightarrow (- \infty, \infty) $ we get
\begin{equation} \langle \chi( \cdot ) \xi( \cdot )| {\cal S} |
\chi( \cdot ) \xi( \cdot ) \rangle = \exp \left( i \int_{-
\infty}^{\infty}  \xi \cdot \dot{\chi} \right)
\end{equation}
 In particular, for a pair of coherent-state histories the decoherence
functional reads
\begin{equation}
d((\xi,\chi)( \cdot ),(\xi',\chi')( \cdot )) = e^{i S[\xi( \cdot
),\chi( \cdot )]- iS^*[\xi'( \cdot ),\chi'( \cdot )]},
\end{equation}
where
\begin{equation}
i S[\xi,\chi] = \langle \xi,\chi| (\frac{d}{dt} - iH)|\xi,\chi \rangle
   \end{equation}
 is the classical phase space action.

\subsection{The generating functional}
\subsubsection{ N - point functions}
A probability theory does not only give probabilities to possible scenaria.
It also provides  expectation values for observables.
In fact, a probability measure can be fully  reconstructed from the knowledge of a
 sufficiently large number of expectation values:
these are known as the moments of the distribution or in physics
as the N-point functions.   We shall write the relevant formulas
in the context of stochastic processes, rather  than single-time
probability theory, for it is the analogue of these expressions
that we shall attempt to generalize in the quantum context.

Let us denote by  $x$  a vector that corresponds to
a point of  a sample
 space $\Omega$ and $\Omega^T$ the space of histories
with elements the paths $x( \cdot )$. Let also  $d \mu(x( \cdot ))$ denote the
  probability measure in the space of paths.
  One then defines the N - point functions
\begin{equation}
G^{(n)}( a_1,t_1; \ldots; a_n,t_n) =  \int d \mu(x( \cdot )) X_{t_1}^{a_1} \ldots
X_{t_n}^{a_n}
\end{equation}
 where
$X^a_t(x( \cdot )) = x^a(t)$ is a function on $\Omega^T$.

The information of the N- point functions is encoded in the generating functional
\begin{equation}
Z[J] = \sum_{n=0}^{\infty} \frac{(i)^n}{n!} \int dt_1 \ldots dt_n \sum_{a_1
\ldots a_n}
G^{(n)} (a_1,t_1;\ldots; a_n,t_n) J_{a_1}(t_1) \ldots J_{a_n}(t_n)  \end{equation}
The generating functional is just the Fourier transform of the stochastic measure
\begin{equation}
Z[J( \cdot )] = \int d \mu(x( \cdot )) \exp( i \int dt X^a_t J_a(t) )
\end{equation}

The $N$-point functions (3.51) fully exhaust the physical content
of the theory; hence the generating functional (3.53) provides a
complete specification of the probability measure. In general, one
can define generating functionals containing less complete
information, e.g. ones that refer to one single observable. For
instance given a function $f$ on $\Omega$ we can define
\begin{equation}
Z_f[J( \cdot )] = \int d \mu(x( \cdot )) e^{i \int dt F_tJ(t)}
\end{equation}
which generates the correlation functions of $f$.
Or more generally, one can define generating
functionals of time-averaged quantities $F$ (functions on $\Omega^T$) as
\begin{equation}
Z_F(j) = \int d \mu(x( \cdot )) e^{ i  F(x( \cdot )) j},
\end{equation}
for some real number $j$.

\subsubsection{The CTP generating functional}

Since the decoherence functional is defined through bounded operators on ${\cal V}$ , its definition can be extended to
 a bilinear functional over all bounded operators on ${\cal V}$: $d: B({\cal V}) \times B({\cal V}) \rightarrow {\bf C}$.

We shall first examine the discrete-time case. Let us consider an
operator $\hat{A}$ on $H$. Then if $A_t$ denotes the corresponding
single-time operator on ${\cal V}$ (see section 2.1), we can see
that
\begin{eqnarray}
d(A_{t_1} \otimes A_{t_2},1) =  \Theta(t_1-t_2)  Tr (\hat{\rho_0} \hat{A}(t_1)
\hat{A} (t_2))  \nonumber \\
+ \Theta(t_2-t_1)
 Tr (\hat{\rho}_0 \hat{A}(t_2) \hat{A}(t_1))
\\
d(1,A_{t_1} \otimes A_{t_2}) =  \Theta(t_2 - t_1)  Tr (\hat{\rho} \hat{A}(t_1)
\hat{A} (t_2)) \nonumber \\
+ \Theta(t_1 - t_2)
 Tr (\hat{\rho} \hat{A}(t_2) \hat{A} (t_1))
\end{eqnarray}
where $\hat{A}(t)$ is the Heisenberg picture operator on $H$:
$e^{i\hat{H}t} \hat{A} e^{-i\hat{H}t}$. The right hand side of
(3.56) and (3.57) are the time-ordered and anti-time-ordered
two-point function for this observable. Similarly we can construct
higher time-ordered and anti - time-ordered functions
respectively, as well as mixed ones, e.g.
$d(A_{t_1},A_{t_2}\otimes A_{t_3})$. They are usually denoted by
$(r,s)$ correlation functions $r$ denoting the number of
time-ordered and $s$ of anti-time-ordered appearances of $A$ in
the expectation value. Such  $N$ - point functions have been first
used in the classic study of quantum Brownian motion by Schwinger
\cite{Schw61}. They are obtained by an object known as the closed
- time - path (CTP) generating functional \cite{Schw61, Kel64}.

If we want to construct an object that
 encodes the information about the $N$-point functions at all
times, we need to go to the continuum limit.

Let us by $A_f$ denote the time averaged version of an operator
$\hat{A}$ on $H$, defined in the way we explained in section 2.3.
Then we define the closed-time-path generating functional
associated to the operator $\hat{A}$ as a function of a pair of
smearing functions $J_+$ and $J_-$ through
\begin{equation}
Z_{\hat{A}}[J_+( \cdot ),J_-( \cdot )] =
d(e^{iA_{J_+}},e^{-iA_{J_-}}).
\end{equation}
The signs $+$ and $-$ correspond respectively to the part that generates time-ordered,
vs anti-time-ordered correlation functions. In general the $(r,s)$ mixed correlation
 function for $A$ will be given by
 \begin{eqnarray}
G_A^{(r,s)} (t_1, \ldots , t_r; t'_1, \ldots, t'_s) \nonumber \\
= (-i)^r i^s \frac{\delta^r}{\delta J_+(t_1) \ldots \delta J_+(t_r)}
 \frac{\delta^s}{\delta J_-(t_1) \ldots \delta J_-(t_s)} Z[J_+,J_-]|_{J_+=J_-=0}.
 \end{eqnarray}

When the Hilbert space carries a representation $U(\chi,\xi) = e^{-iq_{\xi}-i p_\chi}$  of the history Weyl group, there
exist time-averaged versions of the position and momentum operators. We can then construct the
configuration space CTP generating functional as
\begin{equation} Z_q[\xi_+,\xi_-] = d(e^{i(q,\xi_+)}, e^{-i(q,\xi_-)}).
\end{equation}
This generating functional  has been
widely used, mainly because it has a convenient path-integral expression.
One can construct a corresponding effective action through a Legendre transform
of $ W = - i \log Z$ (known as the CTP effective action) \cite{CaHu87}.

But we can also write a generating functional that contains all {\it phase space} correlation
functions. This is simply defined \cite{Ana00} as
\begin{equation}
Z[\xi_+,\chi_+;\xi_-,\chi_-] = d(U(\chi_+,\xi_+),U^{\dagger}(\chi_-,\xi_-)).
\end{equation}
Since our representation of the history group is irreducible, all
physical information about the physical system is contained in the
CTP generating functional (3.61). Indeed, it is the quantum
analogue of the generating functional (3.53) of a general
stochastic process.

\subsection{The Wigner-Weyl transform}
\subsubsection{The canonical case}
In  quantum mechanics a representation $\hat{U}(\chi,\xi)$ of the canonical group enables one to
 construct  a {\it linear} map that  takes
  a large class of Hilbert space operators
to phase space functions. This is known as the Wigner-Weyl transform. It is implemented
as follows:
If $\hat{A}$ is a trace-class operator on $H$ then, we define the function $F_{\hat{A}}(q,p)$ on phase space as
\begin{equation}
F_{\hat{A}}(q,p) = \int d\chi d \xi e^{-i\xi \cdot q - i \chi \cdot p} Tr \left( \hat{U}(\xi, \chi) \hat{A} \right):=
Tr(\hat{\Delta}(q,p)\hat{A}),
\end{equation}
where $\hat{\Delta} (q,p) = \int d\chi d \xi e^{-i\xi \cdot q - i \chi \cdot p} \hat{U}(\chi,\xi)$. This  operator satisfies
\begin{equation}
 \int dq dp \hat{\Delta}(q,p) = \hat{1},
\end{equation}
and its matrix elements in a coherent state basis are given by
\begin{eqnarray}
\langle \chi' \xi'| \Delta(q,p) |\chi \xi \rangle = e^{i\xi \cdot (\chi-\chi') + i q \cdot (p-p') +i (\chi \cdot \xi - \chi' \cdot \xi')} \nonumber \\ \times
\tilde{K} [p + \frac{\xi + \xi'}{2}, q - \frac{\chi+\chi'}{2}],
\end{eqnarray}
in terms of the Fourier transform of the expectation functional
\begin{equation}
\tilde{K}[p,q] = \int d \mu (\chi,\xi) e^{-i\chi \cdot p - i \xi \cdot q} K[\chi, \xi]   .
\end{equation}

Note that by $dq dp$ we denote the standard  Lebesque measure on $\Gamma = {\bf R}^{2n}$,
 normalized by a factor of $(2 \pi)^{-n}$.

This definition can be  extended to bounded operators (at least when the Weyl group is finite
dimensional )
and to a large class of unbounded ones. The Wigner-Weyl transform of a density matrix is known
 as the {\it Wigner function}.
There are two important properties of the Wigner transform
\begin{eqnarray}
\int dq dp F_{\hat{A}}(q,p) = Tr_H \hat{A}, \\ \int dq dp F_{\hat{A}}(q,p) F_{\hat{B}}(q,p) = Tr_H (\hat{A} \hat{B}).
\end{eqnarray}
The operator commutator induces on the phase the Moyal bracket $\{,\}_M$. For a pair of operators $\hat{A}$ and $\hat{B}$ their commutator $\hat{C} = [\hat{A}, \hat{B}]$ is associated with the symbol
\begin{equation}
\frac{1}{i}F_{\hat{C}} =   \{F_{\hat{A}},F_{\hat{B}} \}_M := 2  F_{\hat{A}}\sin\left(\frac{1}{2} \{,\} \right) F_{\hat{B}}
\end{equation}
here $\{,\}$ is the Poisson bracket on phase space, written as a
bilinear operator: $f \{,\} g = \{f,g\}$. The sinus in this
expression refers to its Taylor series viewed as a function of
this bilinear operator.

\subsubsection{The histories analogue}
We can  proceed similarly in the histories case and
to each trace-class operator $A$ on ${\cal V}$ associate a function $F_A$
 on $\Pi$, the space of classical histories as
\begin{equation}
F_A[\gamma] = F_A[q( \cdot ),\xi( \cdot )]  = \int D\xi( \cdot ) D\chi( \cdot ) e^{ - i (q, \xi)(\gamma) - i ( p, \chi)(\gamma) } Tr
(U(\xi,\chi) A).
\end{equation}
This expression is only formal, since the measures $D\chi( \cdot )$ etc do not exist.
What is implied  is $F_A(\gamma) = Tr_{{\cal V}} ( A \Delta(q( \cdot ),p( \cdot )))$. By $\Delta(q( \cdot ),p( \cdot ))$ we denote
a  linear map that is given by
\begin{equation}
\Delta(q( \cdot ),p( \cdot )) = \otimes_{t} \hat{\Delta}(q_t,p_t).
 \end{equation}

If the operator $A$ is a product operator $\otimes_t \hat{A}_t$, then using equation (2.14)
we see that
\begin{equation}
F_{A_f}[q( \cdot ),p( \cdot )] = \exp \left(\int d \mu(t) \log F_{\hat{A}_t}(q_t,p_t) \right)
\end{equation}
It is also easy to calculate the symbol for a time averaged operator $A_f$   by constructing the Weyl transform
  for $e^{iA_fs}$ and expanding around $s=0$. The result is
\begin{equation}
F_{A_f} [q( \cdot ),p( \cdot )] = \int d \mu(t) f(t) F_{\hat{A}}(q_t,p_t)
\end{equation}
Such is for instance the case of position, momentum operators and the Hamiltonian,
so that
\begin{eqnarray}
q_f \rightarrow F_{q_f} = \int d \mu(t) q_t f(t), \\
p_f \rightarrow F_{p_f} = \int d \mu(t) p_t f(t), \\
H_{\kappa} \rightarrow F_{H_{\kappa}} =  \int d \mu(t) \kappa(t) h(q_t,p_t),
\end{eqnarray}
 where $h(q,p) = F_{\hat{h}}(q,p)$ is the Wigner transform of the canonical Hamiltonian.

For more general operators on ${\cal V}$, the Weyl transform is effected by constructing first a suitable discrete-time expression in
 $\otimes_i H_{t_i}$ and then going to the continuum limit.
 It is more convenient to employ the decomposition of the unity for the canonical coherent states in order to compute the trace.
\begin{equation}
 Tr_{{\cal V}} A = \int \prod_{i} d \mu(\chi_{t_i},\xi_{t_i}) \langle \chi_{t_1} \xi_{t_1};\chi_{t_2} \xi_{t_2}
\ldots \chi_{t_n} \xi_{t_n}|A| \chi_{t_1} \xi_{t_1};\chi_{t_2} \xi_{t_2} \ldots \chi_{t_n} \xi_{t_n} \rangle
\end{equation}
For operators that map coherent states into coherent states, the
calculations are easier to perform. Such is, for instance, the
operator $e^{isV}$. We can compute
\begin{equation}
e^{isV} \rightarrow F_{e^{isV}} = \int \prod_t d \mu(\chi_t,p_t) \langle \chi_t \xi_t | \hat{\Delta}(q_t,p_t)| \chi_{t+s} \xi_{t+s} \rangle_{H_t} \end{equation} If we expand this   around $s = 0$ we find that  \begin{equation}
V \rightarrow F_V =  \int dt p_t \dot{q}_t
\end{equation}
where the integral is of a Stieljes type.

Note that equations (3.69) and  (3.76) are, as yet,
 defined for a discretization of the time interval. In order to compute any traces we will always
need to check the finiteness of the expressions at the continuum
limit. We will return to this later in the next subsection
(3.5.3).

\subsubsection{The decoherence functional}
In an analogous manner, one can assign to the decoherence functional a
``function'' on $\Pi \times \Pi$ as
\begin{eqnarray}
W[q( \cdot ),p( \cdot )|q'( \cdot ),p'( \cdot )] =
W[\gamma|\gamma'] = \int D\xi_+( \cdot ) D\chi_+( \cdot ) D\xi_-( \cdot ) D\chi_-( \cdot )
\nonumber \\
e^{-i (q,\xi_+)  - i (p,\chi_+)  + i(q',\xi_-)  + i(p',\chi_-) }
\times  Z[\xi_+,\chi_+;\xi_-,\chi_-]
\end{eqnarray} Given then some operators (these might be  projectors
that correspond to a history proposition) $A$  and $B$ on ${\cal
V}$ we have
 \begin{equation} d(A,B) =  \int D \mu (\gamma) D \mu
(\gamma') W[\gamma| \gamma'] F_A(\gamma) F_B (\gamma')
\end{equation}
where $D \mu(\gamma)$   is a shorthand for $D\chi( \cdot ) D \xi( \cdot )$.

In spite of the general non-definability of the integration measure, there is a very good sense
 in which the $W[\gamma|\gamma']$ exists : as the inductive limit of its discrete-time
 expressions, in complete analogy
with the Kolmogorov's construction of the stochastic probability measure. This proceeds as follows:

In standard quantum mechanics one can define objects that correspond to
classical multi - time probabilities using the Wigner transform \cite{Sri77}. They are of
the form
\begin{equation}
W(q_1,p_1,t_1; \ldots q_n,p_n,t_n) = Tr \left( \hat{\rho}_0 e^{i\hat{H}t_1}
\hat{\Delta}(q_1,p_1) e^{-i\hat{H} t_1} \ldots  e^{i\hat{H}t_n} \hat{\Delta}(q_n,p_n) e^{-i\hat{H} t_n}
\right).
\end{equation}

 These distributions do not define a probability  measure: they are
complex and do not satisfy the Kolmogorov additivity condition.
Rather they are the building blocks of the decoherence functional.
In analogy with the stochastic case if we consider two
discretizations $I = \{ t_1, \ldots t_n \}$ and $I' = \{t_1'
\ldots t'_m \}$ of an interval $T$, we can define the objects
  the
\begin{eqnarray}
W_{n,m}[q_1,p_1,t_1; \ldots q_n,p_n,t_n| q'_1,p'_1,t'_1;\ldots ; q'_m,p'_m,t'_m] = Tr \left( \hat{C}_n^{\dagger} \hat{\rho}_0 \hat{C'}_m \right),
\end{eqnarray}
where
\begin{eqnarray}
\hat{C}_m = e^{i\hat{H}t_1} \hat{\Delta}(q_1,p_1) e^{-i\hat{H}t_1} \ldots e^{i\hat{H}t_m} \hat{\Delta}(q_m,p_m) e^{-i\hat{H}t_m} \end{eqnarray} and similarly for $\hat{C'}_m$.

Let us as  write $\Omega^I$ and $\Omega^{I'}$ the  spaces of
discrete-time phase space histories. They can be equipped with the
standard
 Lebesque measure $\prod_t dq_t dp_t$, so that $W_{n,m}$ can be used to  define
  genuine decoherence functionals $d_{I,I'}$ that satisfy properties (2.3 ).
If we denote  by  $\Omega^T$ the space of phase space histories we can
consider the injection map
$i_{I,I'}: \Omega^n \times \Omega^m \rightarrow \Omega^T \times \Omega^T $.
These maps are {\it measurable}. It is easy to check that the hierarchy of functions $W_{n,m}$ satisfies  an additivity condition
\begin{eqnarray}
 \int dq_{t_1}  dq_{t_2} W_{n,m} [q_1,p_1,t_1; \ldots q_n,p_n,t_n| q'_1,p'_1,t'_1;\ldots ; q'_m,p'_m,t'_m] \nonumber \\ = W_{n-1,m} [q_2,p_2,t_2; \ldots q_n,p_n,t_n| q'_1,p'_1,t'_1;\ldots ; q'_m,p'_m,t'_m]
 \end{eqnarray}
 In complete analogy to Kolmogorov's theorem, the above properties are sufficient to prove the existence of an
 additive, complex-valued, hermitian  measure on $\Pi \times \Pi$, i.e. a decoherence functional $d_{\Pi}$, such that
\begin{equation}
d_{I,I'} = i^*_{I,I'} d_{\Pi}
\end{equation}

It is important to remark that the definition of the decoherence
functional on phase space took place with respect to the {\it
measurable subsets } of $\Pi$, which define a Boolean algebra.
This is  clearly distinct from the logic of projectors on the
Hilbert space ${\cal V}$. This is what enabled us to sidestep the
non-definability of a decoherence functional from the
discrete-time expressions.

This construction does not highlight the general structure of the decoherence functional. To see this, it
is necessary to compute the Wigner transformations of the operators ${\cal S} $ and ${\cal U}$.

When the Hamiltonian is quadratic, the coherent states are
Gaussians and the calculation of traces reduces to Gaussian
integrals. In this case the functional relations of operators is
preserved by the Weyl - Wigner transform. For the harmonic
oscillator, we get
\begin{eqnarray}
{\cal U} \rightarrow F_{{\cal
U}}&=& e^{ - i H_{\kappa}} , \kappa(t) = t \\ {\cal S} \rightarrow
F_{{\cal S}} &=&  \exp \left( -\frac{1}{2}  [ \omega (q_{t_f} -
q_{t_0})^2 + \omega^{-1} (p_{t_f} - p_{t_0})^2 + i ( p_{t_f} \cdot
q_{t_0} - q_{t_f} \cdot p_{t_0})] \right) \nonumber \\ \times
e^{i/2 \int_{t_i}^{t_f} dt (p_t \cdot \dot{q}_t - q_t \cdot
\dot{p}_t)}
 \end{eqnarray}
 In the case of more general
Hamiltonians the calculations are more difficult to perform. But
if we assume that the interval upon which histories are defined is
the whole real line, the boundary condition forces that
\begin{equation}
F_{{\cal S}}= e^{i \int dt p_t \dot{q}_t}
\end{equation}   The operator ${\cal U}$  is unitary,
hence a transformation
 $ A \rightarrow {\cal U} A {\cal U}^{\dagger}$  preserves the trace.
 The trace is also preserved by the Weyl - Wigner transform ,
 hence on phase space ${\cal U}$ corresponds to a trace-preserving
 automorphism ${\bf T}$ of the algebra of functions on the space $\Pi$ of phase space paths.
 Explicitly this would be the continuum limit of
\begin{equation}
{\bf T} = T_{t_1} \otimes \ldots \otimes T_{t_n} \end{equation}
where $T_t$ corresponds to the automorphism of the algebra of
single - time functions $A \rightarrow T_t[A]  $ given by the
Moyal bracket  version of the Heisenberg equations of motion
\begin{equation}
 \frac{\partial}{\partial_t} T_t[A] = \{H,T_t[A] \}_M
 \end{equation}
with $T_0[A] = A$.

In general the decoherence functional for phase space paths in a time interval
$[t_i,t_f]$
 will read
\begin{equation}
d(A,B) = \int_{\Gamma_{t_i} \times \Gamma_{t_f}}  d x
\lambda_x^*(F_A)\lambda_x(F_B) \end{equation}
where by \\
- $x$ we denote points on the boundaries $\Gamma_{t_i} \times \Gamma_{t_f}$. It is then a collective index
for $(q_{t_i},p_{t_i},q_{t_f},p_{t_f})$.
It is obtained by the Weyl - Wigner  transform of the boundary operator with
respect to the
$rs$ indices in equation (2.19).
\\
- $d x = dq_{t_0} dp_{t_0} dq_{t_f} dp_{t_f}$ is the standard measure
 on $\Gamma_{t_0} \times \Gamma_{t_f} $ . \\
- $\lambda_x( \cdot )$ is a family of complex valued measures on
the space of paths that have a functional dependence on boundary
points $x$ which incorporates  the actual initial state of the
system. If by ${\bf T}$ we denote the automorphism generated by
${\cal U} $ then
\begin{equation}
\lambda_x(A) = \int d \mu(\gamma) F_{{\cal A}^x{\cal S}}(\gamma) {\bf T}(A)(\gamma)
\end{equation}
In this equation $F_{{\cal A}^x{\cal S}}$ is the Weyl symbol associated with the operator ${\cal A}^x{\cal S}$ of equation (2.23). The colective variable $x$ again corresponds to the Weyl-Wigner transform of the indices $(r,s)$ of equation (2.19).

In this expression it is very clear that phases appear in the probability
assignment solely because of the geometric phase encoded in the
 operator ${\cal S}$. This has been argued in \cite{AnSav00}, but in the present context it is clearer, since
the automorphism ${\bf T}$ makes no reference to complex numbers
in its definition. The presence of complex numbers in the
decoherence functional is purely due to the presence of a $U(1)$
connection on phase space, as encoded in the function $F_{{\cal
S}}$.

\subsection{The stochastic  limit}
Rather than considering the decoherence condition (1.5) as a law of nature, that has to be exactly satisfied
(as the consistent histories interpretation does), we can view it
 as a condition for the approximation of the physical system by a classical probabilistic
 theory.
We remarked how the unequal time   pseudo - probability
distributions $W_{n,m}$ do not satisfy the Kolmogorov additivity
conditions. Perhaps a smeared version of them would
(approximately) satisfy  them so that one would get decoherence.
So one can try to define smeared pseudo - probability
distributions like
\begin{eqnarray}
\bar{W}_{n,0}(\bar{q}_1,\bar{p}_{1},t_1; \ldots \bar{q}_n,\bar{p}_n,t_n)
\nonumber \\
= \int dq_1 d p_1 \ldots d q_n dp_n \chi_{\bar{q}_1 \bar{p}_1}(
q_1, p_1) \ldots \chi_{\bar{q}_n \bar{p}_n}( q_n, p_n)
W_{n,0}(q_1,p_1,t_1;\ldots; q_n,p_n,t_n)
\end{eqnarray}
Here $\chi_{\bar{\chi} \bar{\xi}}$ denotes a smeared characteristic function of a cell
centered around $\bar{\chi} \bar{\xi}$. This will depend on some
parameters
$V$ which will determine the volume of the cell, within which smearing is
effected.

The objects $\bar{W}_{n,0}$ are expectation values. They can be
properly normalized if we divide them with the  smearing volume.
In that case they can be taken as the discrete-time probability
densities that might correspond to a  measure. If these smeared
densities satisfy the Kolmogorov criterion, (which is to be
expected in many systems given sufficient smearing) they would
define an  classical probability measure, that would give an
effective stochastic description for the quantum system.

\subsubsection{General operators}
The above description is valid for general observables and not
only the generators of the canonical group. Indeed if $\hat{A}$ is
a self-adjoint operator with continuous spectrum $\Sigma$, one
defines its corresponding generating functional
$Z_{\hat{A}}[f_+,f_-]$  as in equation (3.58). Now if $x \in {\bf
R}$ denote points of the spectrum of $A$, we can construct a
decoherence functional in the space of histories $x( \cdot ): T
\rightarrow \Sigma$ by an analogous expression to (3.78)
\begin{equation}
W[x( \cdot )|x'( \cdot )] = \int Df_+( \cdot ) Df_-( \cdot ) e^{-i (x,f_+) + i (x',f_-) } Z_{\hat{A}}[f_+,f_-]
\end{equation}
For any two functions $F$ and $G$ on the space of paths, we will have
\begin{equation}
d(F,G) = \int Dx( \cdot ) Dx'( \cdot ) F[x( \cdot )] G[x'( \cdot )] W[x( \cdot )|x'( \cdot )]
\end{equation}
The distribution $W$ can again be defined as the inductive limit of the discrete-time distributions
$W_{n,m} (x_1,t_1;\ldots;x_n,t_n|x'_1,t'_1 ; \ldots ; x'_n,t'_n)$ as in equation (3.81), but with the operators
\begin{equation}
\hat{\Delta}(x) = \int dJ e^{-ixJ} e^{i\hat{A} J}
\end{equation}
substituting $\hat{\Delta}(p,q)$.

Again one can look for the classical limit by constructing smeared characteristic
functions $\chi_{\bar{x}}(x)$
for subsets of $\Sigma$.
  It is convenient to use a Gaussian function
for $\chi_{\bar{x}}$. For instance
\begin{eqnarray} \chi_{\bar{x}  } (x) = \exp \left( - \frac{1}{2 \sqrt{V}}
  ( x - \bar{x})^2  \right)
\end{eqnarray}

\subsubsection{Smearing}
Let us give a description of how the above prescription for
finding the classical limit. We may start with discrete time
histories with  $n$ time -steps, which we shall simply call $t$
 and
consider the smearing functions $\chi_{\bar{x}}$.
\begin{equation}
\chi_{\bar{x}( \cdot )} = \prod_t \chi_{\bar{x}_t} = \exp \left( - \frac{1}{2
\sqrt{V}} \sum_t (x_t - \bar{x}_t)^2 \right)
\end{equation}

Then we evaluate  the decoherence functional  at a pair of $\chi$
's (actually their corresponding positive operators)
 to be
\begin{eqnarray}
 d(\chi_{\bar{x}( \cdot )}, \chi_{\bar{x}'( \cdot )}) = \nonumber \\
\int D J_+ D J_- Z_{\hat{A}} [J_+,
J_-] \left( \int Dx Dx' \chi_{\bar{x}( \cdot )} [x( \cdot )] \chi_{\bar{x'}( \cdot )} [x'( \cdot )] e^{ -
i (J_+,x) + i (J_-, x')} \right) \nonumber \\ = V^n \int D J_+ D J_- Z_{\hat{A}}[J_+/\sqrt{V}, J_-/\sqrt{V}]
\nonumber \\
\times
\exp
\left( - \frac{1}{4}
(J_+,  J_+) - \frac{1}{4} (J_-, J_-) - i( \bar{x}, J_+ ) + i
(\bar{x}', J_-)
 \right) \hspace{2cm}
\end{eqnarray}
By $(J,J') $ we imply here a discrete sum $\sum_t J_t J'_t$. When, we go to the continuous limit it will imply $\int d \mu(t) J_t J'_t$.

\subsubsection{ The probability measure}
Assume now that with sufficient coarse - graining  we can get approximate
satisfaction of the decoherence
condition for disjoint $\chi_{\bar{x}( \cdot )}$ and $\chi_{\bar{x}'( \cdot )}$.
The next step is to assume that the probabilities $ p(\chi_{\bar{x}}) = d ( \chi_{\bar{x}}, \chi_{\bar{x}}) $ can be used to define a probability measure
\begin{equation}
p[\bar{x}( \cdot )] = \frac{1}{V^n}  d ( \chi_{\bar{x}}, \chi_{\bar{x}} )
\end{equation}
 This is  standard practice. It   corresponds to the mathematical operation of
extending a classical probability measure  that is
 defined in only a part of the lattice of propositions
  (in this case  a  semi-lattice), to
the whole of the lattice. This gives then a  generating functional
(note that $p[\bar{x}( \cdot )] $ has no multiplicative dependence
on $V^n$ hence it is safe to go to the continuous limit)
\begin{eqnarray}
Z_A[J] = \int D\bar{x} p[\bar{x}( \cdot )] e^{i (\bar{x},J)} = \nonumber \\
  \int D J_+ \exp \left( - \frac{1}{2}
 (J_+,J_+)  -  \frac{1}{2} |J - J_+|^2 \right)
\nonumber \\
\times
Z[(J_+)/\sqrt{V}, (J -
 J_+)/\sqrt{V}]
\end{eqnarray}
This is the generating functional of a stochastic
 process for a classical observable $A$, that is obtained as the classical limit of a
 general quantum mechanical operator $\hat{A}$.

The above construction can be repeated for phase space observables with no modifications.
 In this case, the representation of the canonical group, provides a {\it natural metric on phase space}, which can be used in order to construct smearing functions. In this case, a parameter analogous to $V$ plays the role of the volume of the phase space cell (with respect to this metric), within which one smears.

Details on how to obtain the classical stochastic limit of
quantum systems with this method, together with a number of
examples, are found in \cite{Ana00g}.

\subsection{Summary}

After giving a brief review of the canonical group construction
and the histories version of classical mechanics, we showed how to
construct a large class of representations of the history group,
using coherent state techniques.  A particular nice result was,
that for well-behaved quantum field theories the representation of
the canonical group uniquely determines  one for the history
group.

We then showed how to encode the correlation functions for generic
observables of the theory into a CTP generating functional. The
Wigner-Weyl transform offered a way of representing quantum
 mechanical objects on the phase space and
 define a continuous-time decoherence functional as the continuous limit
 of discrete-time ones.

Finally, we developed a general procedure for taking the classical
probability limit of quantum mechanical histories.

\section{Discussion}
We shall now discuss a number of topics that explain or  put into
context the results of the previous two sections.

\subsection{Time averaging}
First, we need to address a rather important issue, that we left
uncommented. What is the role of the parameter $\tau$ that enters
 the definition of the  time integral? It appeared there
originally in order to render the measure dimensionless, so that
operators $A_f$ would be dimensionally the same with their
canonical counterparts $\hat{A}$.

In the case where $T$ is compact, we remarked that $\tau$ can be
chosen as to normalize the measure to unity. But in the more
general case, that $T = {\bf R}$, this cannot be done and one
would have to accept $\tau$ as an
 additional  parameter entering the histories quantum theory. On one hand,
  it would not appear into the physical predictions of
 the theory: the values of the decoherence functional are independent of $\tau$.
 Nonetheless, it would be present n the definition of the time averaged operators
and perhaps in the physical correspondence with classical
observables.

One possible idea is to substitute  all integrals over $d \mu(t)$
with the limit as $ \tau \rightarrow \infty $ of
$\frac{1}{\tau}\int_{-\tau/2}^{\tau/2} dt$. Classical quantities
of the form
\begin{equation}
q_f = \lim_{\tau \rightarrow \infty} \frac{1}{\tau} \int_{- \tau/2}^{\tau/2} dt q_t f(t)
\end{equation}
are more naturally interpreted as time-averaged values of the observable $q$. This implies that we can enlarge the space of possible
test-functions. It would suffice to demand, that the smearing functions
$f$ are  {\em constant} outside a compact set (rather than zero), in order for the  integral to be defined.

This could have as immediate consequence, that  the boundary
Hilbert spaces at infinities would not  be one-dimensional, as is
the case of when $f$ is of compact support.

But this would severely weaken our uniqueness theorem. We need to
have a unique translationary invariant ``vacuum'' vector, in order
for the uniqueness theorem to hold. This is not any more true, if
$f$ is not a function of compact support: any vector $|z(\cdot)
\rangle$
 with
 constant values of $z(t) $ would be translationary invariant.

 The representation theory for this history group would
    therefore be very different; in fact, the history group itself
 is different. Intuitively, one  expects that the representation we  would obtain
 from such a construction  would be a reducible one:
a direct integral of representations like the ones we constructed,
each labelled  by  different boundary conditions for the coherent
states as $t \rightarrow \pm \infty$.

 These considerations will be taken further in another paper.

\subsection{The decoherence functional}
We tried various different ways to define a continuous-time
decoherence functional. The straightforward analogy with
Kolmogorov's construction failed, because we cannot continuously
embed the lattice of single-time propositions to the lattice of
history ones. We were then left with two choices:
 one is to incorporate the information about the initial condition in an object that is
  extended in time, rather than a density matrix as is the case in
   the canonical approach. This might be operationally meaningful (after all the initial state corresponds
to a preparation that takes place in a time interval), but it contradicts our intuition that information about the system can be encoded at a single moment of time, without any need of knowing anything about its past history. (In a sense, such a construction might be
considered as the violation of the analogue of the Markov condition for stochastic processes.)

The other alternative, is to define the decoherence functional with respect to the structure of propositions about phase space histories.
 This involves abandoning continuity, but in phase space the natural condition is measurability and using this we can construct a mathematically sensible continuous-time decoherence functional. Operationally it is a very satisfactory construction: phase space measurements exhaust the physical content of quantum theories.
 But one might raise the objection that we sacrificed the quantum logic structure of history propositions in order to achieve this.

This objection is valid, assuming one considers quantum logic to
be a fundamental part of quantum theory. This is however an
interpretational attitude towards quantum theory, which we do not
feel obliged to adopt. But even should we concede this point,
 we  could still argue, that the true quantum logic is the
one corresponding to time-averaged history propositions,
and the standard  single-time one just an approximation.

 It is nonetheless true, that our construction would be conceptually more complete,
and aesthetically more satisfying,
 if we were able to provide  a reconstruction theorem: that the knowledge of  the
 decoherence functional on phase space,
allows us to  uniquely construct the Hilbert space of the theory,
the decoherence functional defined as a bilinear functional on the histories Hilbert
 space  and perhaps get some
 correspondence between phase space symmetries and quantum mechanical unitary operators.
 This would be an analogue of Wightman's
 reconstruction theorem in quantum field theory \cite{SW64}: constructing  the Hilbert
space,  the vacuum and the representations of
 the history group from the correlation functions. The analogy is very accurate,
  because the decoherence functional on phase space is equivalent
to the CTP generating functional and thus incorporates information about all
correlation functions.

So far, we have not been able to find a direct way to prove such a
theorem. Of course, one could always proceed indirectly: define
the Wightman functions from the CTP generating functional, from
them the canonical Hilbert space, the vacuum and the Hamiltonian,
and then repeat the construction of section 3  to construct the
history Hilbert space and the representation of the history group.
Even though this lends plausibility in the existence of a
reconstruction theorem, it does not provide any physical or
mathematical insight on the structure of history  theories.

\subsection{The classical limit}
 The identification of the history Hilbert space was based on the representations
  of the history group. When we have a representation of a
group, we inherit all structures associated to it: coherent states,
their symbols and the Weyl - Wigner transform. The phase space, then,
 appears as the most fundamental ingredient of the quantum theory.

Indeed, through the Weyl-Wigner transform we can cast
 quantum mechanical histories in a language that makes only indirect references
 to a Hilbert space and is completely based on classical phase space objects.
 This is important, because on phase space we know how to
 implement coarse-grainings, that are of interest for a wide class of physical systems.

For instance, in many particle systems, one could study
coarse-grainings of the Boltzmann type (focusing on a description
in terms of densities on a single-particle phase space) and derive
their stochastic behavior, by the method described in 3.6. This
might provide a way to proceed towards a declared aim of the
consistent histories scheme: to find how hydrodynamic variables
and their quasi-deterministic evolution laws  arise from quantum
theory \cite{GeHa9093, BrHa96, Hal9899, BrHa99, CaHu99}. In fact,
all types of coarse-graining of classical statistical mechanics
can be implemented for phase space histories.

Another area, where our results are relevant is in the study of
back-reaction of quantum fields on geometry. The semiclassical
treatment assumes that we can couple the Einstein tensor to the
expectation value of a quantum stress-energy tensor.

For quantum fields in curved spacetime, the stress-energy tensor
is not defined as an operator or even an operator-valued
distribution on the Hilbert space of the theory. This is why it
has to be renormalized \cite{Wald94}, but, even so, one cannot
remove the divergences from its correlation functions. Our
construction suggests that one could first take the stochastic
limit for the field in a histories version of the theory, and then
construct a {\it classical} stress-energy tensor from the
classicalized field. This would give a fully consistent scheme for
dealing with the back-reaction of the matter to geometry, without
the dangerous assumptions involved in computing expectation values
of stress-energy tensors. This idea has been tentatively developed
in \cite{Ana00g}.

\subsection{Perturbation theory}

In practice, we cannot explicitly construct the Hilbert space of
the theory and the basic objects for most interesting physical
systems. That is, why we rely on approximation methods, like
perturbation theory. In analogy to quantum field theory, we could
perhaps develop a perturbation expansion for the decoherence
functional, together with a renormalization scheme, in order to
adequately treat non-linear systems

The first problem we would face, is the generic inadequacy of
perturbation theory to deal with real-time evolution. In this case
we have to expand the operator $e^{-i \hat{H}t}$ in powers of the
coupling constant,
 something that becomes increasingly inaccurate with large values of $t$.  In standard quantum theory,
this problem is addressed by performing the perturbation expansion, not to the evolution operator, but to
 its resolvent $(E- \hat{H})^{-1} $, which is essentially its Fourier transform.

The CTP generating functional plays the same role, since it is a ``Fourier transform''
 of the decoherence functional. In the CTP formalism, the perturbation theory is well defined -for instance,
in the path integral representation - and its accuracy does not
depend on the time $t$. This leads to a perturbative evaluation of
the decoherence functional, that does not suffer from the problems
of real-time perturbation theory. As such, it provides a valuable
tool for the construction of powerful  approximation schemes in
the histories programme.

\subsection{The histories quantization programme}
The main motivation of this paper is to be found in the histories
quantization programme. This  aims to exploit
 the covariant nature and  the richer content  of the histories approach, in order to study  quantum theories
of systems with non-trivial temporal structure. The eventual aim is a theory of quantum gravity.

So far the programme has dealt with quantum fields in curved spacetime
 \cite{Ana00, Nol00}(where, unlike the canonical case,  we can construct a
theory accepting an instantaneous Hamiltonian) and with
constrained systems. The two laws of time transformation have
enabled a treatment of parameterized systems \cite{SavAn00}
(prototypes of general relativity), in which the problem of time
does not appear. More recent results involve the more elaborate presence of Poincar\'e groups in quantum field theory \cite{Sav01a} and 
the appearance of a representation of the spacetime diffeomorphism group in the histories version of general relativity \cite{Sav01b}.

The main obstacle to further generalisation has been the restriction to Fock representations for the history group, and hence only to
 quadratic systems. In this paper, we have constructed
 a larger class of representations  and therefore enlarged the domain of
 applicability of the programme. We have also indicated, how a perturbative construction of history theories could
  be implemented.

This will provide tools for continuation of the programme: it will be possible to rigorously construct covariant quantum theories for a
large class of systems, at least with the same level of rigor as the canonical approach.

\begin{appendix}

\end{appendix}

\end{document}